\pdfoutput=1
\documentclass[preprint,authoryear,review,12pt]{elsarticle}

\usepackage{graphicx}
\usepackage{epsfig}

\usepackage{amssymb}

\newcommand{\eg}{{\it e.g.},\ }
\newcommand{\ie}{{\it i.e.},\ }
\newcommand{\etal}{{\textit{et al}}.}

\journal{Icarus}

\begin{document}

\begin{frontmatter}



\title{Limits on the Size and Orbit Distribution \\ of Main Belt Comets}


\author{S. Sonnett}

\author{J. Kleyna}
\author{R. Jedicke}

\address{\textit{Institute for Astronomy, University of Hawaii at Manoa, 2680 Woodlawn Drive, Honolulu, HI, 96822}}

\author{J. Masiero}
\address{\textit{Jet Propulsion Laboratory, M/S 321-520, 4800 Oak Grove Drive, Pasadena, CA, 91109}}

\begin{abstract}
The first of a new class of objects now known as main belt comets
(MBCs) or ``activated asteroids'' was identified in 1996.  The seven
known members of this class have orbital characteristics of main belt
asteroids yet exhibit dust ejection like comets.  In order to
constrain their physical and orbital properties we searched the
Thousand Asteroid Light Curve Survey \citep[TALCS;][]{mas09} for
additional candidates using two diagnostics: tail and coma detection.
This was the most sensitive MBC survey effort to date, extending the
search from MBCs with $H \sim 18$ (D $\sim$ 1 km) to MBCs as small as $H \sim 21$ ($D \sim 150$ m).

We fit each of the 924 objects detected by TALCS to a PSF model
incorporating both a coma and nuclear component to measure the
fractional contribution of the coma to the total surface brightness.
We determined the significance of the coma detection using the same
algorithm on a sample of null detections of comparable magnitude and
rate of motion.  We did not identify any MBC candidates with this
technique to a sensitivity limit on the order of cometary mass loss rate of
about 0.1 kg/s.

Our tail detection algorithm relied on identifying statistically
significant flux in a segmented annulus around the candidate object.
We show that the technique can detect tail activity throughout the
asteroid belt to the level of the currently known MBCs.  Although we
did not identify any MBC candidates with this technique, we find a
statistically significant detection of faint activity in the
entire ensemble of TALCS asteroids.  This suggests that many main belt
asteroids are active at very low levels.

Our null detection of MBCs allows us to set 90\% upper confidence
limits on the number distribution of MBCs as a function of absolute
magnitude, semimajor axis, eccentricity, and inclination.  There are
$\lesssim$400000 MBCs in the main belt brighter than $H_V=21$
($\sim150$-m in diameter) and the MBC:MBA ratio is $\lesssim$1:400.

We further comment on the ability of observations to meaningfully
constrain the snow line's location.  Under some reasonable and simple
assumptions we claim 85\% confidence that the contemporary snow line
lies beyond 2.5 AU.

\end{abstract}


\end{frontmatter}


\section{Introduction} 

The classical view of comets as icy conglomerations and asteroids as
chunks of rock has been supplanted in the last decade by the
realization that a population of objects exists between the two
traditional extremes --- comets are dirty iceballs and asteroids are
icy dirtballs with the relative contribution and morphological
structure of the ice and rock giving rise to the classical views of each object.  
Thus, in the classical view, asteroids always 
exhibit bare nuclei photometric properties \citep{jew08} while comets are 
characterized by a transient atmosphere and/or tail \citep{jew09}.  The
early compositional distinction between asteroids and comets was further supported
by the apparent bimodality in their orbit distributions --- the known
asteroids had nearly circular orbits confined to the torus of objects
between Mars and Jupiter while the known comets had highly eccentric
orbits taking them beyond Jupiter, Neptune and even out to the Oort
Cloud\footnote{Asteroids and comets may be distinguished by their Tisserand 
parameter with respect to Jupiter ($T_J$) - comets typically have $T_J < 3.0$ 
while asteroids generally have $T_J > 3.0$.}.  The modern view was ushered in by 
the discovery of comet-asteroid transition objects.  In the ensemble of 
known small bodies there exist comets with asteroidal dynamical 
properties \citep[\eg][]{cha96,fer05}, asteroids on cometary orbits
\citep{bin93,fer05}, and Damocloids \citep*[inactive comets with
  dynamical properties of Halley-family and long-period
  comets;][]{jew05}.

This work focuses on those objects that appear to be comets --- their
morphology being consistent with a cometary nature in the sense that
they exhibit comae or tails --- but have asteroid-like orbits embedded
in the main belt of asteroids.  To this end, we have performed a
well-characterized search for low-level cometary activity amongst a
sample of nearly 1000 asteroids fromTALCS \citep{mas09}.

The first cometary main belt object, now known as a main belt comet
(MBC) or an ``activated asteroid'', was identified in observations of
what is now known as Comet 133P/Elst-Pizarro \citep[hereafter,
  EP;][]{els96}.  It exhibits recurrent dust ejection over several
weeks or months \citep{boe96}.  \citet{hsi06b} described a detailed
study of EP and showed that its cometary activity is correlated with
heliocentric distance. \citet{hsi10} then showed that EP's activity is
also seasonal --- there is a region on the surface that becomes active
when it experiences its local `summer'.  They then explored two
dynamical scenarios that might explain EP's transient cometary
activity and its orbital characteristics embedded in the outer region
of the main belt.

Their first scenario is that EP began as a Jupiter family comet (JFC)
but migrated inward via both non-gravitational (\ie cometary
outgassing) and gravitational influences.  However, none of the
simulations of the dynamical evolution of JFC test particles under
solely gravitational influences result  in an
inclination as low as EP's \citep{fer02}.  \citet{ipa07} showed that
non-gravitational forces can be strong enough to bring EP to its
contemporary orbit, but its current activity level is unlikely to have
produced enough of a perturbation to do so \citep{hsi06b}.

Alternatively, \citet{hsi06b} suggested that EP could be the first of
a large number of asteroids containing a reservoir of ice beneath
their surface.  This scenario is consistent with thermal evolution
models of large asteroids that escaped primordial heating and with
evidence of aqueous alterations in meteorites
(\citealt{gri89,fan89,pri09}).  Furthermore, \citet{pri09} show that
crystalline water ice from the time of their formation can survive
within the interior of outer main belt asteroids.  The MBCs were not
discovered earlier because they are rare and their activity is both
weak and/or transient.  The release of buried volatiles requires some
triggering event like an impact (even a small one) or the warming
effect of perihelion passage (\citealt{jon90,sco05,hsi04,cap10}), and
their detection requires regular monitoring of a large sample of
asteroids.  These conditions were not met until the advent of modern
wide-field asteroid surveys.

The origin of EP and other MBCs is of interest in planetary formation
in part because they offer an opportunity to identify the location of
the `snow line' --- the heliocentric distance at which ices condensed
in the early solar system (\eg \citealt{sas00,ken08}).  MBCs should
only be found outside the snowline assuming that there has been only
limited heliocentric mixing of the asteroids.  Furthermore, the
existence and properties of a large sample of MBCs will provide tests
of asteroid thermal models, common origin scenarios, and dispersion
mechanisms.  For example, considering significant instead 
of limited heliocentric mixing, MBC studies can provide hints on how 
this mechanism functioned in the early solar system environment.  

In an attempt to identify more MBCs when only two were known,
\citet{hsi06a} conducted a targeted survey of about 300 asteroids in
the outer main belt and found one more in the orbital region of the
other two MBCs.  All three have similar orbit characteristics with
$3.156 \leq a \leq 3.196$ AU, $0.165 \leq e \leq 0.253$, and
$0.24^{\circ} \leq i \leq 1.39^{\circ}$ (Table \ref{tab:knownmbcs}).  Two of
the three belong to the Themis dynamical family while the third
(P/Read) has an eccentricity slightly higher than the Themis family's
upper limit.  \citet{nes08} suggest that P/Read used to be a Themis family 
member that is dynamically evolving away from its parent cluster.  
\citet{hsi06b} estimated that the 
MBC:MBA ratio is $\sim$1:300 and measured MBC mass loss rates in 
the range of $0.01 - 1.5$ kg/s compared to typical cometary mass loss 
rates of $\sim 10^{-3} \lesssim \dot{M}$ (kg/s) $\lesssim 10^{3}$
\citep[\eg][]{lam04}.  Since then, four additional MBCs have been
discovered outside the vicinity of the original three MBCs with two of
them located in the middle belt.  Some characteristics of all seven
objects are provided in Table \ref{tab:knownmbcs}. For at least two of the
MBCs in Table \ref{tab:knownmbcs}, the observed activity is 
most likely the result of asteroid collisions and be devoid of recurrent
activity characteristic of comets 
\citep[596 Scheila and P/2010 A2/Linear;][]{jew10,sno10,jew11}.

One of the major difficulties in searching for new MBCs and mapping
their distribution is detecting and quantifying their subtle cometary
nature.  Several techniques have been employed in the past to identify
low activity comets; most were designed to search for faint comae.

One method is to identify optical emission lines of typical cometary
gases \citep[\eg][]{coc86}.  Detecting these faint spectroscopic features 
(\eg the CN(0-0) and C$_2$ bands at 388nm and 517nm, respectively)
requires high $S/N$ objects and a relatively large amount of telescope
time for each object making it difficult to employ on a large sample
of main belt candidates.

Another method to search for cometary activity requires multiple
photometric observations of the targets over a wide range of phase
angles to identify non-asteroidal photometric behavior.  The flux of
scattered light off an asteroidal target is proportional to the
product of its phase function, the inverse square of its heliocentric
distance, and the inverse square of its geocentric distance.  Failure
of the target's photometric profile to follow this behavior suggests a
variable coma \citep{har90}.

Speckle interferometry has also been used to distinguish comae but it
is limited to only the brightest objects with $m_{V} \leq 14$,
limiting this process to the largest $\sim$28,000 asteroids or
$\sim$7\% of the known main belt objects (\citealt{dru89,bow07}).  The
process also requires considerable telescope time for each target.

\citet{gil09} attempted to identify main-belt comets morphologically
using the expected FWHM-broadening of the target PSF perpendicular to
the direction of motion.  In a follow-up paper, \citet{gil10}
identified one object that may be either an MBC candidate or a regular comet
 and set an upper limit on the number of
MBCs in the main belt of $40 \pm 18$ to a limiting size of $\sim
1$~km (absolute magnitude $H \sim 16$).

To identify MBC candidates \citet{luu92} compared MBC candidates to
stellar profiles with the expectation that a wider asteroidal profile
would indicate the presence of coma.  We adopted and refined this
method in our search for new MBCs amongst the asteroids identified in
TALCS \citep{mas09}.  That survey identified 924 asteroids with multiple 
$S/N>5$ detections of each object 
using the CFHT's MegaPrime camera.  Our goal was to
carefully examine each TALCS asteroid for low-level cometary activity
and determine the MBC number distribution as a function of semi-major
axis ($a$), eccentricity ($e$), inclination ($i$), and absolute
magnitude ($H$) or diameter ($D$).  We used two techniques to identify
cometary activity around otherwise asteroidal objects: one for tail
detection and another for comae detection.  We then corrected for
observational selection effects and derived limits on the unbiased
orbit and number distribution of the MBCs.

\section{Observations}

We obtained a large set of asteroid detections from TALCS
\citep{mas09} which was designed to measure light curve properties of
$\sim 1000$ Main Belt asteroids with diameters in the range 0.5~km $<
D <$ 10~km.  The survey was conducted with the Canada-France-Hawaii
Telescope's MegaPrime camera whose image plane is instrumented with an
array of 36 CCDs that each contain $2048 \times 4612$ pixels.  With a
pixel scale of $0.185^{\prime\prime}$/pixel MegaCam covers a field of $\sim
1^{\circ} \times1^{\circ}\;$.  TALCS used the $g'$ and $r'$
filters with integration times of 20 and 40 seconds yielding
5-$\sigma$ detections at about 23.3 and 24.3 magnitudes respectively \citep{fuk96}.

\begin{figure}
\begin{center}
\includegraphics[width=5.5in,height=3.7in]{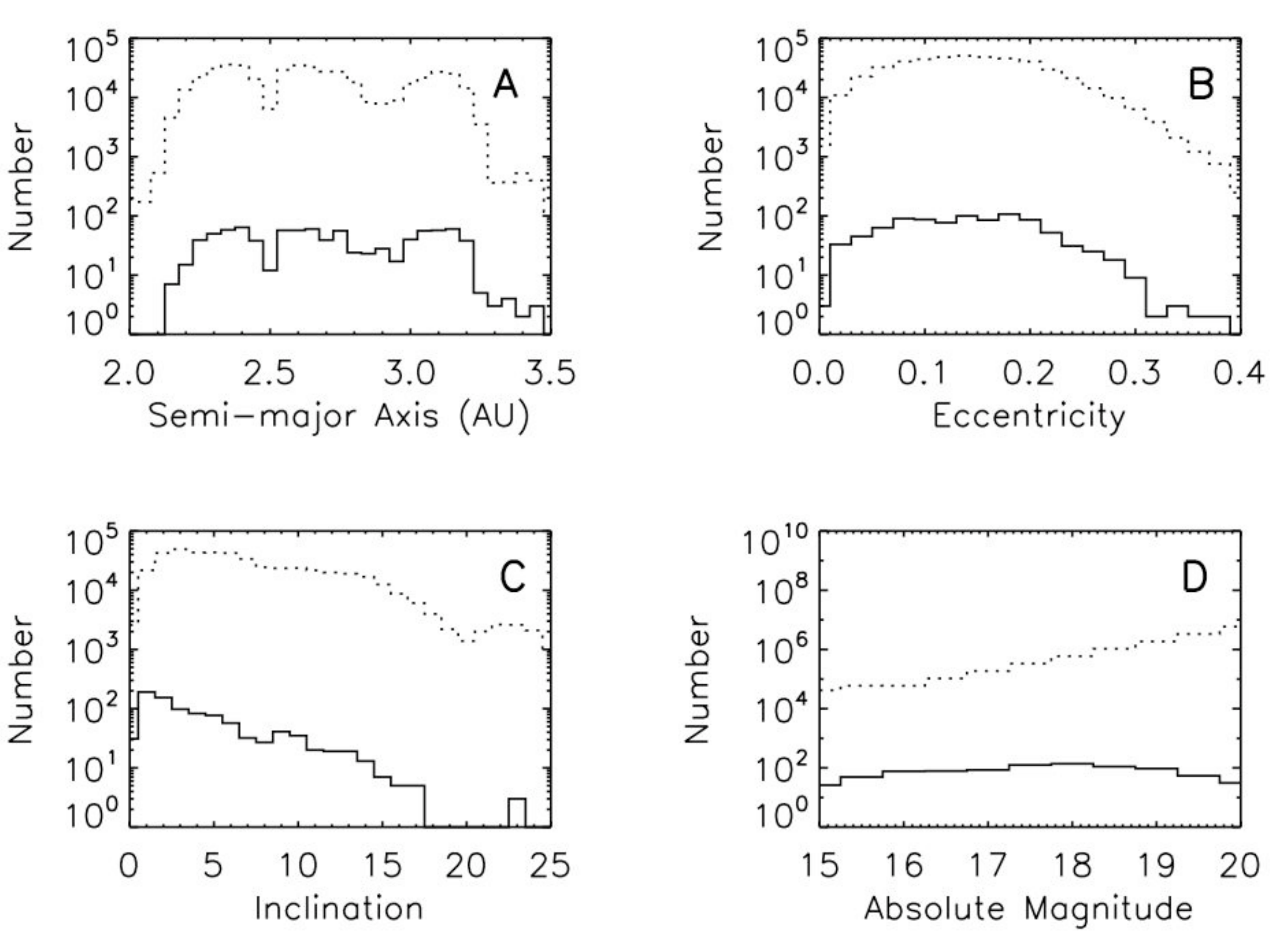}
\caption{ Number distributions of TALCS objects \citep[solid
    line,][]{mas09} and those in the ASTORB database \citep[dotted
    line,][]{bow07} as a function of (A) semimajor axis, (B)
  eccentricity, (C) inclination, and (D) absolute magnitude.}
\label{fig:dist_TALCS_vs_AST}
\end{center}
\end{figure}

Figure \ref{fig:dist_TALCS_vs_AST} shows that the number distribution
of TALCS objects is well-sampled through the main belt.  In the inner
belt ($a < 2.50$~AU), middle belt ($2.501 < a < 2.824$~AU) and outer
belt ($a > 2.824$~AU) there were 286, 287 and 352 objects
respectively.  It is unsurprising that the distribution of TALCS
objects in semi-major axis and eccentricity matches that of the known
objects because in these elements the TALCS survey is similar to most
other asteroid surveys.  TALCS is biased against high-inclination
objects (see Figure \ref{fig:dist_TALCS_vs_AST}C) because it surveyed
a relatively small region on the ecliptic for a short period of time.
The median $H \sim 18$ corresponds to $\sim 1$ km in diameter
(\citealt{bow07}).  This size range is well suited to a MBC search based
on the known objects listed in Table \ref{tab:knownmbcs}.

\section{The Search for Tails}
\label{s.SearchForCometaryTails}

The three MBCs observed by \citet{hsi06a} had tails or dust trails but
weak or nonexistent comae.  This observation motivated us to develop
an algorithm to identify MBC tails but the problem is complicated by
the facts that they 1) are much fainter than typical cometary tails,
2) are transient and may appear or disappear during the course of the
TALCS survey and 3) may appear at any position angle and change their
orientation from night-to-night.  Our tail detection algorithm needed
to be robust against all these possibilities.

\subsection{Method}

\begin{figure}
\includegraphics[width=5.5in,height=5.5in]{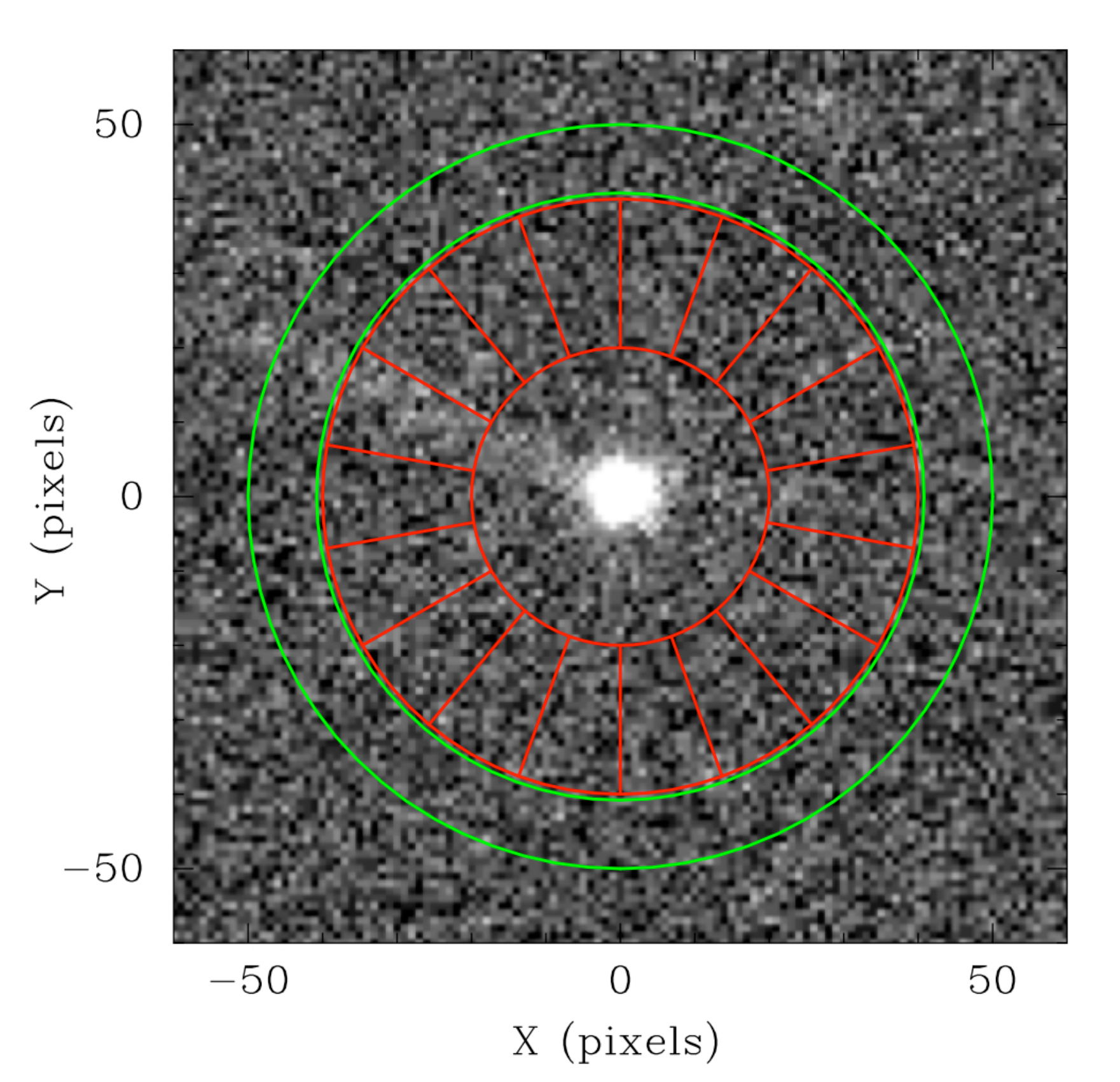}
\caption{\label{fig:elstpizpie} Our scheme for detecting tails or
  trails around asteroids superimposed on an image of Elst-Pizarro.  The red
  detection annulus is subdivided into 18 polar segments and a tail is
  detected by recording the brightest segment (here, at approximately
  10 o'clock).  The green annulus is used to compute a median
  background which is subtracted before the detection procedure.}
\end{figure}

Our tail identification strategy was to divide an annulus of sky
around each detection into eighteen $20^\circ$ truncated pie segments
(Figure \ref{fig:elstpizpie}) and search for an anomalously bright
segment, comparing the result to a set of comparison stars.  Using 
segments is preferable to summing the light in the
entire annulus because the $S/N$ of a tail detection increases as the
square root of the number of segments as long as the tail falls into
only one segment. The benefit of noise suppression dominates even
though the number of opportunities for a false detection scales with
the number of segments. This technique is an analog to a traditional
matched detection kernel using a detection region that mimics the
shape of the detected object.

We used a detection annulus extending from $4''$ to $8''$ from the
asteroid.  For an asteroid at a geocentric distance of $\sim 2$ AU, this 
is equivalent $\sim 6000-12000\;$km from the nucleus.  To reduce contamination of the MBC
candidate by other faint astronomical sources in the image we rejected
asteroid and comparison star detections with a neighboring object
within $11''$. A larger annulus increases the $S/N$ only modestly but
greatly increases the number of images rejected because of
neighboring objects.  The diameter of the inner edge of the annulus
was selected to avoid most of the light from the target's PSF.
Trailing is not an issue because the inner radius of the detection
aperture is much larger than the typical trailing distance of
$\lesssim 0.3''$.

For each image (detection) $i$ of each asteroid we determine the flux
of the brightest annular segment and repeat the procedure for 
nearby stars in the same image with fluxes similar to the asteroid.
Next, we rank the brightest segment of the asteroid among the
brightest segment of the stars with a cumulative parameter $f_i$.
For example, $f_i=0.1$ would mean that the brightest segment for
asteroid detection $i$ is in the top 10\% of all the brightest
segments of the comparison stars. Under the null hypothesis of no MBC
activity the values of $f_i$ are uniformly distributed between 0 and 1.
In the presence of MBC activity there is an excess of detections
having small values of $f_i$.

A compelling feature of this method is that the null hypothesis
distribution of $f_i$ is well defined, non-parametric, and immune to
many types of systematics.  The set of $f_i$ that are used to combine
data across observing nights are independent of variations in
observing conditions and concentrations of background contamination
(\eg unresolved galaxies) because it is calibrated with stars observed
under the same conditions within the same image.

\subsection{Sensitivity}


The sensitivity of any search method depends on the strength of the
signal (in our case, the tail) relative to random background noise and
systematic background artifacts.  To test the sensitivity of our tail
identification method we performed a Monte Carlo simulation using
simplified Gaussian statistics to generate a set of $f_i$ with signal
strengths expressed as a fraction of the standard deviation of counts
over the entire annulus.  We then compared the set to the uniform null
hypothesis.  We assumed 50 identical observations per object and 100
calibration stars.  In the actual data, we have $38\pm 17$ valid
measurements of each asteroid, and each measurement uses $54\pm 19$
calibration stars within $\pm 1$ magnitude of its object.

We repeated the procedure for $N_{\rm{seg}}=$ 1, 9, 18, and 36 segments
corresponding to angular widths of $360^{\circ}$, $40^{\circ}$,
$20^{\circ}$, and $10^{\circ}$, respectively.  For each $N_{\rm{seg}}$ we
conducted 1000 trials and computed the median Kolmogorov-Smirnov
\footnote{The Kolmogorov-Smirnov test is a standard non--parametric
  statistical test comparing two distributions based on the maximum
  difference between their cumulative distributions.  A low
  probability $p$ is derived if it is unlikely that the two
  distributions are drawn from the same underlying distribution.}
probability with which a tail of the given strength is recovered. For
comparison, we also computed the recovery strength for simple additive
and median stacking of the images assuming no systematic noise and
perfect tail alignment among images.

\begin{figure}
\includegraphics[width=5.5in,height=5.5in]{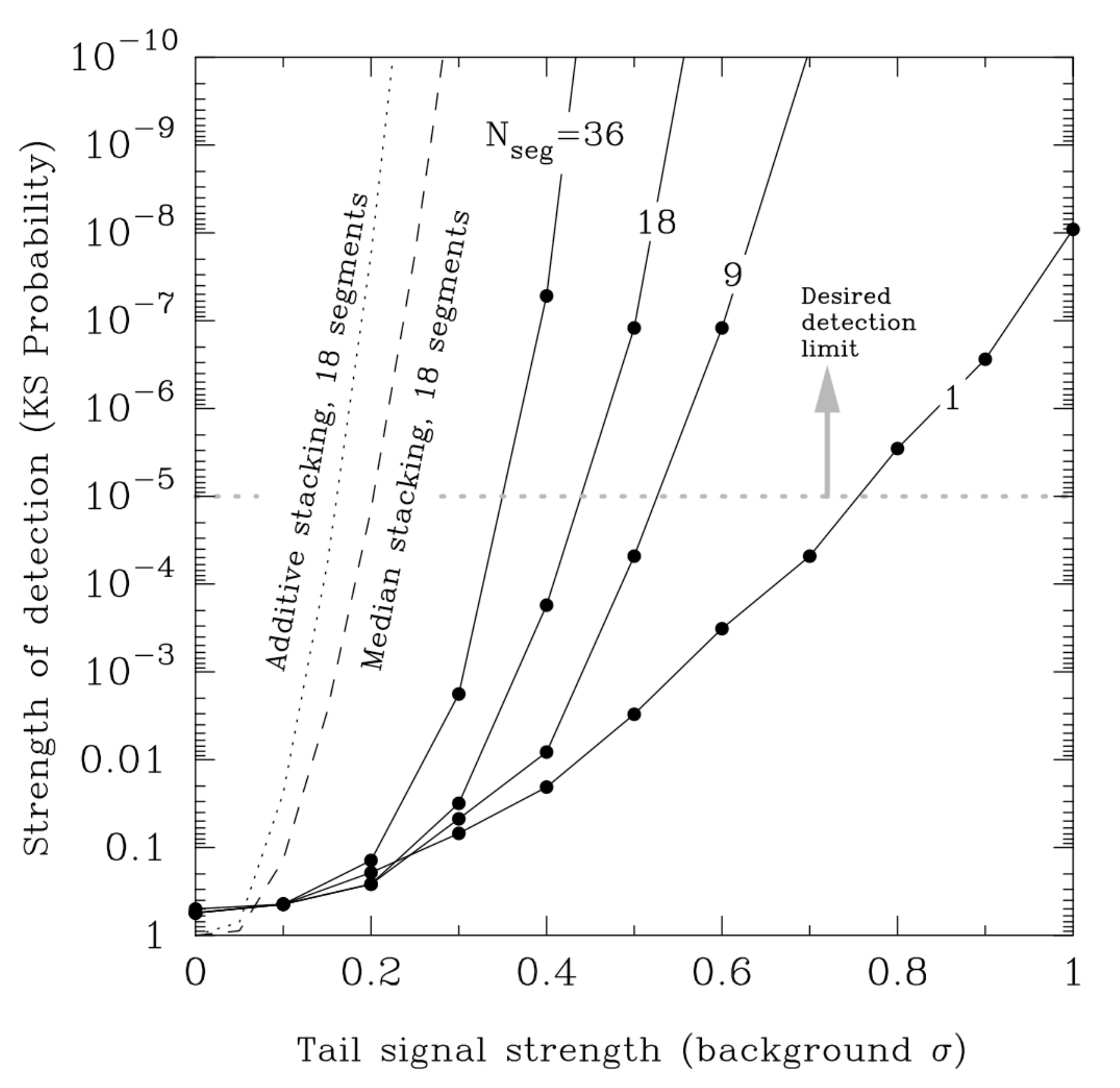}
\caption{\label{fig:slicemethodstrength} Sensitivity of our segmented
  annulus tail search method for MBCs.  We assume a detection annulus
  containing Gaussian noise with a known standard deviation and
  express the signal strength of the tail (horizontal axis) as a
  fraction of the whole-annulus noise. The technique provides the tail
  recovery strength (vertical axis; the Kolmogorov Smirnov
  probability) for each signal strength for the given number of
  segments $N_{\rm{seg}}$.  We also show the recovery strength that
  would be obtained using additive and median image stacking assuming
  perfect tail alignment and no contamination.}
\end{figure}

Figure \ref{fig:slicemethodstrength} shows that our 18 segment scheme
can detect tails at the $p = 10^{-5}$ significance if they have a tail S/N of 
$\gtrsim 0.45$.  Increasing $N_{\rm{seg}}$ allows the
detection of fainter tails but we chose $N_{\rm{seg}}=$18 as a
compromise between sensitivity and the ability to detect wide tails.

We selected a $10^{-5}$ significance as our threshold detection level
because a given asteroid will achieve this level by chance only 1 time
out of $10^5$, so our $\sim 1000$ asteroid sample has a 1 in 100
chance of containing a value this large.  In other words, a signal at
this level for an asteroid in our sample would provide roughly 100 to
1 evidence for MBC activity if we believe that there is a good chance
that there is one MBC in our sample.

Additive or median stacking of the images followed by selecting the
brightest segment can detect fainter tails than our approach but these
techniques are sensitive to image artifacts as well as tail rotation
and transience.  Even though
additive stacking is the most sensitive method in
Fig. \ref{fig:slicemethodstrength}, it would not suppress image
artifacts and we do not consider it a viable option.  Relative to
median stacking our method trades a factor of two in sensitivity in
the ability to connect images obtained under different observing
conditions.  It permits the tail to rotate in position angle between
images and allows for the possiblity that activity ceases in some
images (in which case median stacking could lose the signal entirely).

We tested the sensitivity of our method on real data using MBC images
obtained with the University of Hawaii (UH) 2.2-m telescope
(\citealt{hsi06a}).  For the three known MBCs there are 17 to 25 300s
exposures totaling 1.5 to 2.1 hours of exposure time --- the
equivalent of forty minutes cumulative exposure on the larger CFHT
telescope used for TALCS.  Hence, the total exposure time for the
three MBCs is similar to that of a CFHT TALCS object imaged on a
hundred 30s exposures.  The shorter TALCS/CFHT exposure, with four
times less signal than the 2.2-m data, will suppress the $S/N$ of the
brightest segment by a factor of 2 in any individual exposure relative
to the UH 2.2-m data.  Thus, the UH 2.2-m data set allows the
identification of tails that are half as bright as the CFHT/TALCS data
but the larger number of exposures in TALCS allows a more robust
rejection of the null hypothesis uniform $f$ distribution.

\begin{figure}
\includegraphics[width=5.5in,height=5.5in]{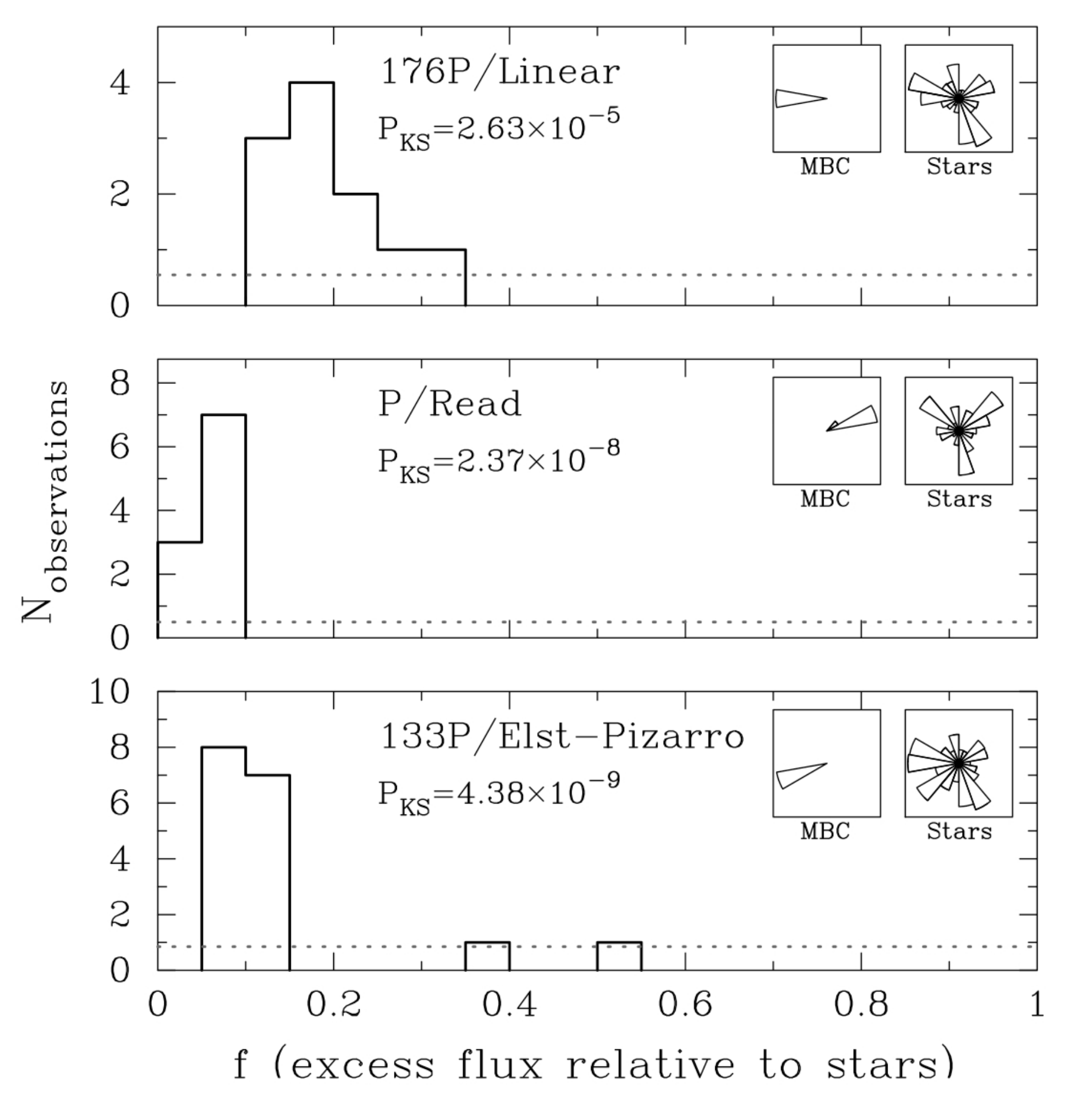}
\caption{\label{fig:88mbcstats} Histogram of the $f$ parameter of
  excess light relative to field stars (see text) for three known MBCs
  observed by \citet{hsi06b}.  The flat dotted line gives the
  null-hypothesis flat distribution.  In each case, the
  Kolmogorov-Smirnov probability $P_{\rm KS}$ strongly rules out the
  null hypothesis.  The inset boxes are radial histograms depicting
  the direction of the brightest polar segment for the MBC (left box)
  and for the calibration stars (right box). The radial length of a
  polar bin is proportional to the number of exposures in which this
  bin corresponded to the brightest direction.}
\end{figure}

Figure \ref{fig:88mbcstats} shows the $f$ distributions produced by
our method for three known MBCs.  The number of entries in the
histograms are considerably smaller than the number of exposures
because our stringent background rejection criteria removed many
images from the sample.  In each case, the Kolmogorov-Smirnov
probability $p_{\rm KS}$ is much smaller than unity, strongly ruling
out the null hypothesis. Even the faintest MBC, 176P/Linear, for which
a tail is invisible to the eye in individual exposures, shows a tail
detection in each frame using our polar segment method. Moreover, from
the inset circular histograms it is evident that the brightest segment
consistently points in the same direction providing a second signature
of MBC activity.

Table \ref{tab:knownmbcstails} provides two measures of the
technique's sensitivity for three known MBC tails: in terms of 1) the
tail flux as a fraction of the object's central flux and 2) the $g'$
magnitude within the brightest detection segment.  We used the
published $R$ magnitudes for the three MBCs \citep{hsi06a} and the
color transformations of \citet{jes05} to calculate $g'$.  Two of the
MBCs have a total tail magnitude in the detection aperture of
$g\approx 23$ while the third is about one magnitude brighter.  Thus,
the technique is capable of detecting tail activity corresponding to a
few percent of the brightness of the object itself.

\begin{figure}
\includegraphics[height=2.5in,angle=-90]{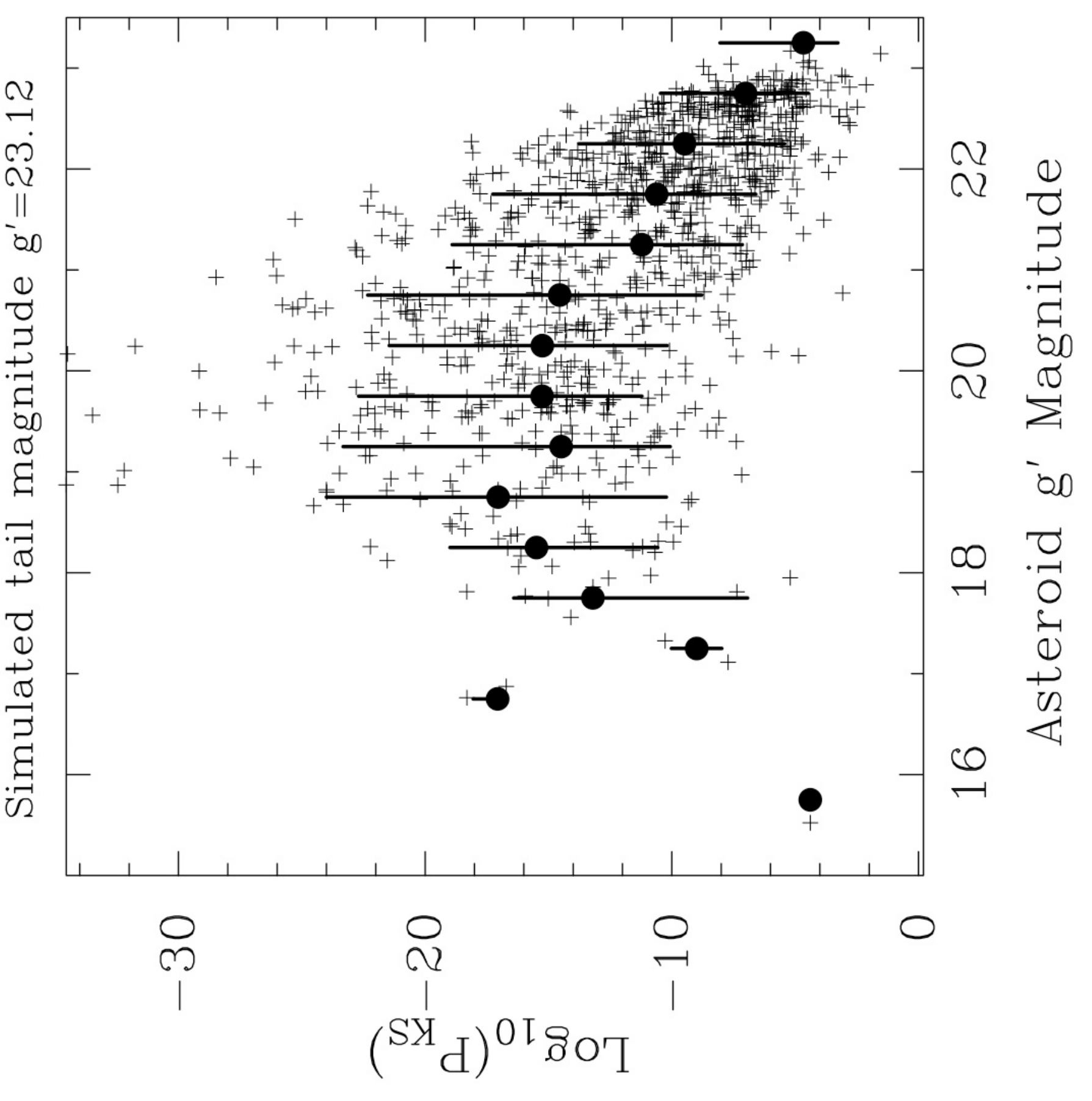}
\qquad
\includegraphics[height=2.5in,angle=-90]{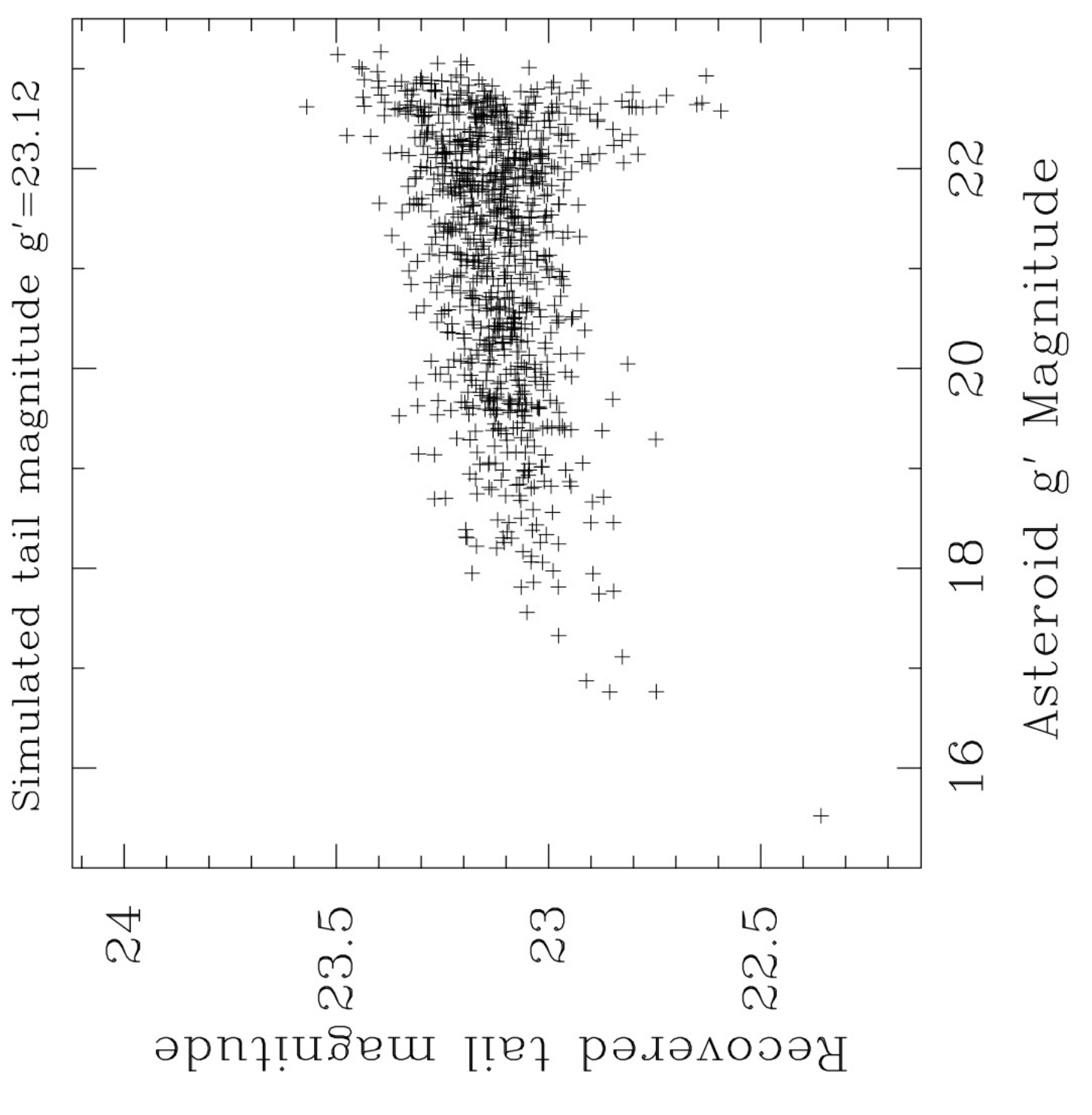}\\
\vskip10pt
\includegraphics[height=2.5in,angle=-90]{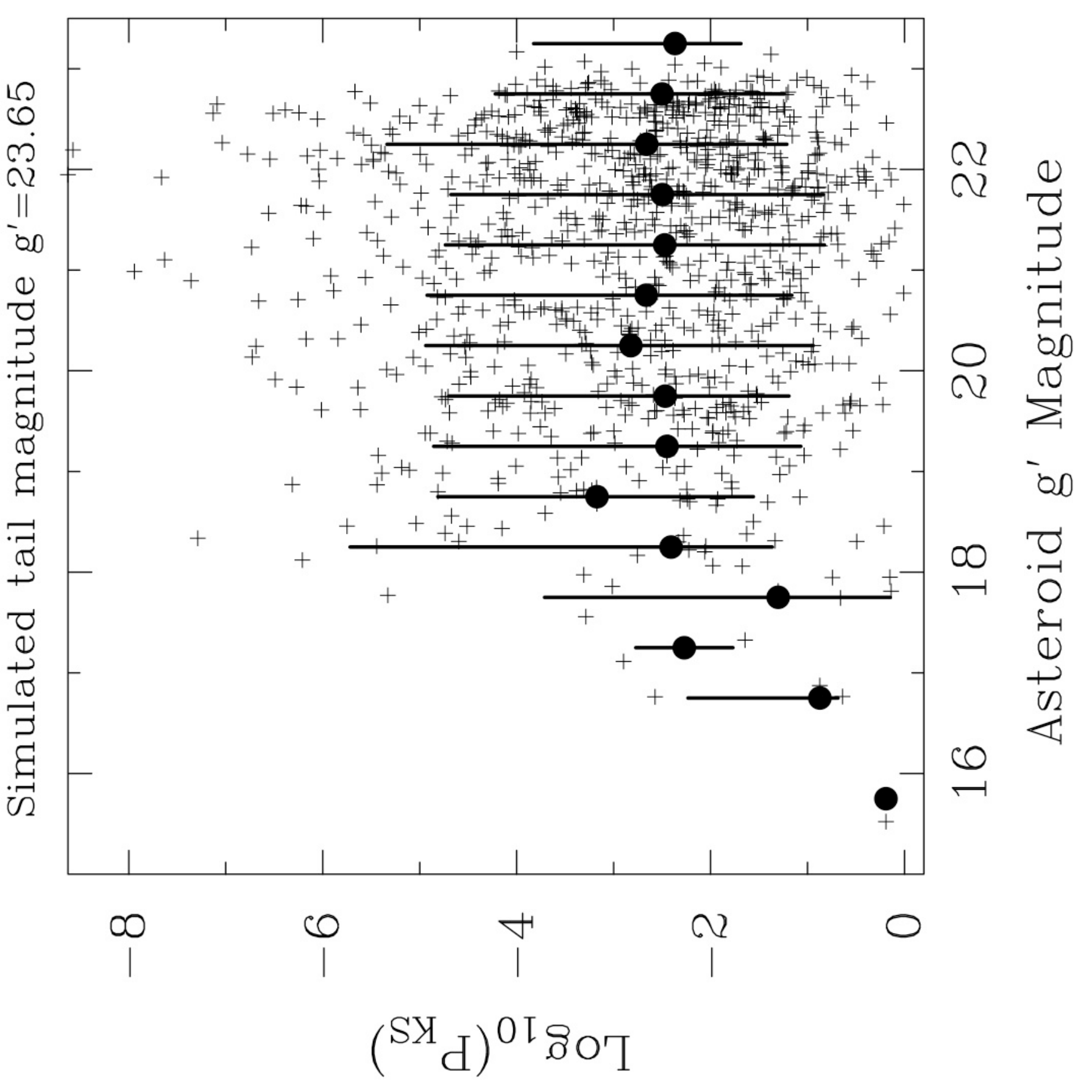}
\qquad
\includegraphics[height=2.5in,angle=-90]{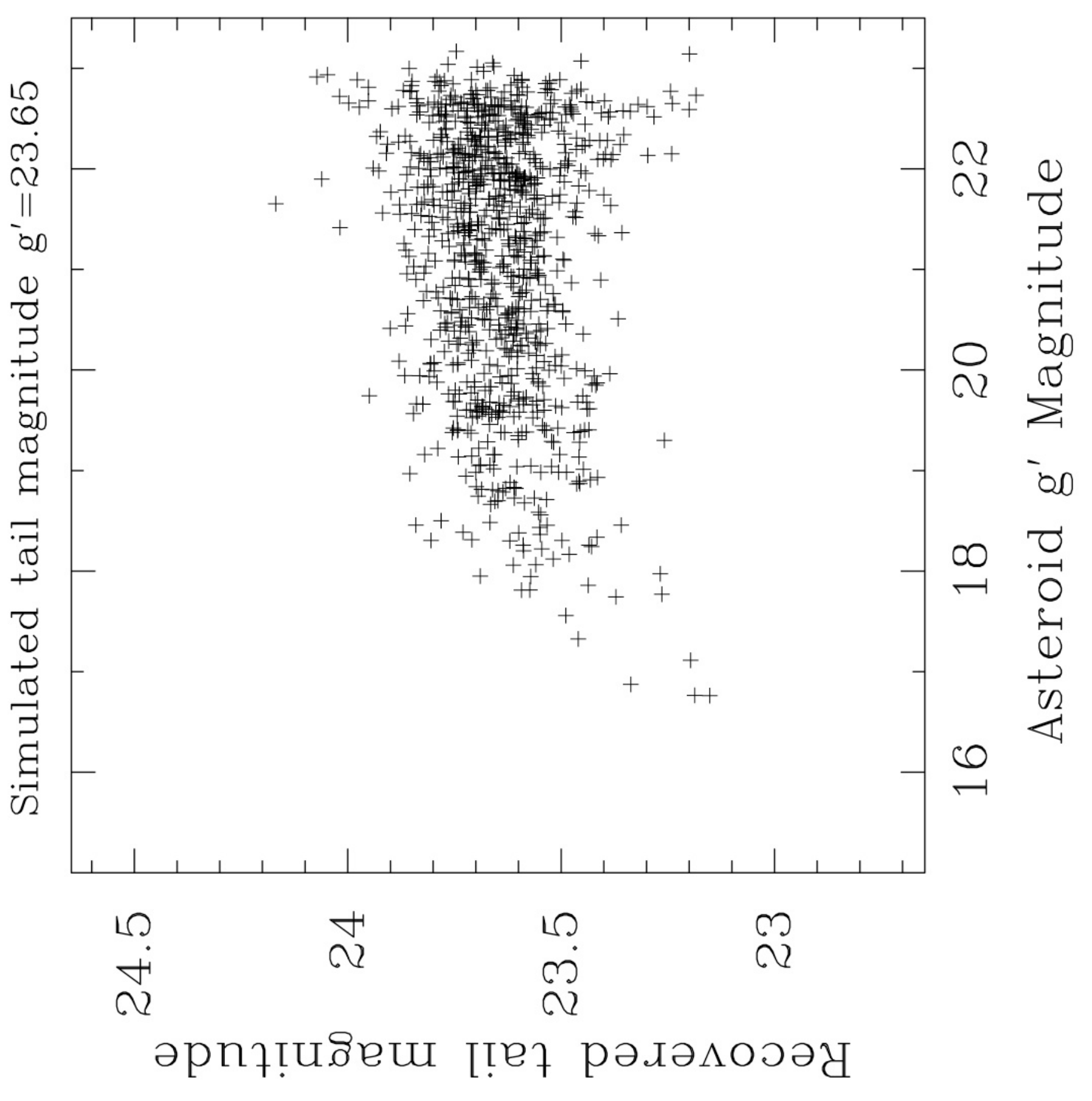}

\caption{\label{fig:talcssimtail} Results of our tail detection
  algorithm on TALCS data when we added an artificial tail of a fixed
  magnitude to each asteroid detection.  The left column of figures
  shows the Kolmogorov--Smirnov probability with which a tail is
  detected; the large point is the median in each half--magnitude bin
  and the bar is the region containing 90\% of points.  The right
  column is the derived magnitude of the recovered tail.  The top row
  of figures provides the results for a simulated tail with a total detection aperture
  magnitude of $g^{\prime} = 23.12$, slightly fainter than the faintest known
  MBC.  The bottom row shows the result for a tail 0.5 magnitude
  fainter (below our detection limit).}
\end{figure}

Finally, we measured our ability to recover artificial tails inserted
into the TALCS data. Figure \ref{fig:talcssimtail} shows that tails
with $g' =23.12$ in the detection slice, comparable to the tail brightness of the fainter
known MBCs, would almost always be recovered with $p_{\rm KS}\ll
10^{-10}$.  Since any detection with $p_{\rm KS}< 10^{-5}$ in our
sample of $\sim 10^3$ objects is indicative of genuine MBC activity at
the 99\% confidence level, this technique is clearly sensitive to MBC
activity at the level of the known objects.  Figure
\ref{fig:talcssimtail}C shows that tails that are only 0.5 magnitudes
fainter than 23.12 would not be reliably detected.  Nevertheless, the presence of
fainter tails is detectable over the ensemble as a whole and their
magnitude is recovered correctly as shown in Figure
\ref{fig:talcssimtail}D.

\section{The Search for Comae}
\label{s.SearchForCometaryComae}

Comets may exhibit a coma despite having a weak or undetectable tail
\citep[\eg 49P/Arend-Rigaux,][]{mil88} but detecting the contribution
of faint coma to the nuclear PSF is difficult.  Thus, we developed a
technique to identify faint comae by expanding upon the work of
\citet{luu92}.  For each TALCS object $n$, we fit a stacked image of
all detections for the object ($F_n$) to a linear combination of a
target-specific asteroid PSF model ($F_{A,n}$) and target-specific
isotropic coma PSF model ($F_{C,n}$):
\begin{equation}
F_n(i,j) = f_a\: F_{A,n}(i,j) + f_c\: F_{C,n}(i,j) \; .
\label{eq.FnFaFc} 
\end{equation}
\noindent where ($i,j$) is the pixel in the stacked image.  Modeling
the target flux is necessarily imprecise because 1) the PSF varies from
night-to-night and across the field-of-view of the wide-field CFHT
MegaPrime camera, 2) asteroids move during the course of the exposures
producing trails of different lengths for each object and even for the
same object in different images because they were taken up to two
weeks apart, and 3) not all comae are isotropic.

The mass loss rate ($\dot{M}$) from a comet is given by \citep{luu92}
\begin{equation}
 \frac{\dot{M}}{{\rm kg/s}} = 10^{-3}\; \pi\; \eta \;
 \frac{\bar{a}}{{\rm \mu m}} \; 
 \frac{\rho}{{\rm kg \; m^{-3}}} \; 
 \biggl (\frac{r}{{\rm km}} \biggr)^2 \;\;
 \frac{{\rm arcsec}}{\phi} \; 
 \frac{{\rm AU}}{\Delta} \;
 \sqrt{\frac{{\rm AU}}{R}} \;\;,
\label{eq:eq.dMdt}
\end{equation}
\noindent where $\rho$ is the grain density, $\bar{a}$ is the average expelled 
grain radius, $\eta$ is the ratio of the flux density of the coma to that of the
nucleus, $r$ is the radius of the target, $\phi$ is the photometric diaphragm, 
and $R$ and $\Delta$ are the heliocentric and geocentric distances, respectively.  
We expect the particles ejected from MBCs are small - similar to cometary material
with grain radii of 0.5 $\mu$m, and we expect that they have densities typical of 
solid rocky material \citep{bri02,luu92}.  We therefore used 0.5 $\mu$m for grain radius 
and 3000 kg/m$^{3}$ for grain density in our calculations.  From eq. \ref{eq.FnFaFc}, 
$\eta\equiv\frac{f_c}{f_a}\;$, and thus we can determine the mass loss rate for the 
TALCS objects.

We calculated each objects's heliocentric and geocentric distances
from their orbits and estimate their radii using \citep{lam04}:
\begin{displaymath}
r \sim \frac{673 \times 10^{-H/5}}{\sqrt{A}}\;\;,
\end{displaymath}
\noindent where $A$ is the geometric albedo.  While TALCS did obtain 
photometry in two different filters for most of the targets the $S/N$ was not
sufficient to measure accurate colors for most objects (from which we
could have assumed a taxonomic type and albedo).  Therefore, we
resorted to assigning a heliocentric distance-dependent albedo: $A =
0.134 / 0.103 / 0.076$ for objects in the inner/middle/outer main belt
bounded by 2.064~AU$<a\le$2.501~AU, 2.501~AU$<a\le$2.824~AU, and
2.824~AU$<a\le$3.277~AU, respectively (\citealt{jed98,kla92}.

\subsection{Method}

\begin{figure}
\begin{center}
\includegraphics[width=7.0in, height=3.5in,angle=90]{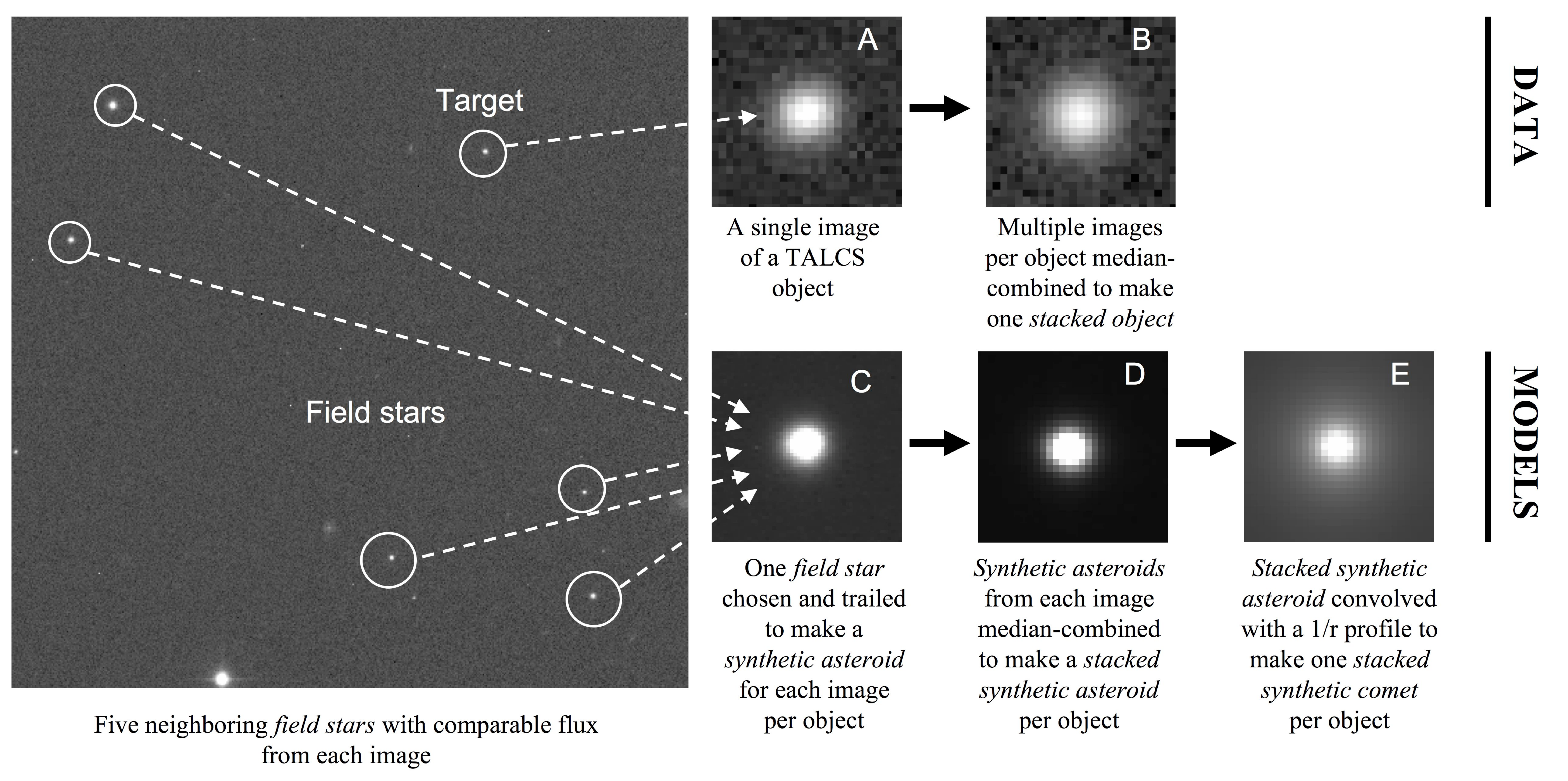}
\caption{\label{fig:sampleimages} Schematic representation of the
  production of stacked object images and their accompanying synthetic
  asteroid and coma models: (A) a single image of a TALCS object; (B) a
  median-combined image of multiple detections of the same object to
  produce one stacked object; (C) one field star chosen based upon its
  similarity to the target in flux, then trailed to make one synthetic
  asteroid for each image per object; (D) synthetic asteroids
  median-combined to make one stacked synthetic asteroid per object;
  (E) each stacked synthetic asteroid is convolved with a $1/r$
  profile to make a stacked synthetic comet (coma-only) per object}
\end{center}
\end{figure}

Figure \ref{fig:sampleimages} provides a schematic representation of
how we produced the three components of our linear fits in
eq. \ref{eq.FnFaFc}: 1) the stacked target object images,
$F_{n}(i,j)$, to which we fit our models, 2) the synthetic asteroid
models specific to each object, $F_{A,n}(i,j)$, that incorporate the
same PSF, trailing and stacking as the target objects, and 3) the
synthetic coma models, $F_{C,n}(i,j)$, that, again, incorporate the
same PSF, trailing and stacking as the target objects.

\subsubsection{Constructing the Stacked Target Image: $F_{n}(i,j)$}


We began by extracting 200 $\times$ 200 pixel ($37^{\prime\prime} \times
37^{\prime\prime}$) thumbnail images (hereafter ``thumbnails'') for each target
object $n$ from each image $m$.  The thumbnail is large enough to
encompass background and a broad coma profile but also small enough to
exclude most field stars.  We median-stacked thumbnails of each
background-subtracted flux-normalized object from images with $\leq
0.8^{\prime\prime}$ seeing as determined by nearby stellar profiles.  The
stacking was performed with sub-pixel offsets when centroiding the
objects.  We combined $g^{\prime}$ and $r^{\prime}$ images
because our concerns are with a morphology that would
manifest itself similarly in both bands.  The background was assumed
to be the median of all pixels in the thumbnail excluding those within
$3^{\prime\prime}$ \hspace{0.0cm} of the target's center and was subtracted from
each raw image.  Since each target object appeared in multiple images
under different seeing conditions and moving at slightly different
rates each night the stacked object images had complicated PSFs.

\subsubsection{Constructing the asteroid model: $F_{A,n}(i,j)$}

We retrieved thumbnails for five nearby bright but unsaturated stars
($10,000 - 50,000$ ADU) for each object in each image (the median
object flux being $\sim2,500$ ADU).  We then background-subtracted,
flux-normalized, and median-combined the field star thumbnails in a
fashion parallel to \S 4.1.1 to produce stacked star images
$F_{S,n,m}$ to be used as PSF models for point sources specific to each TALCS
object $n$ in each image $m$.  Constructing models from nearby field
stars in this way maximized the similarities between the model and
target PSFs' morphological properties.  Then, since the asteroids
moved during each $\sim 30$ second exposure at rates that may have
changed slightly from night to night, we artificially trailed the
stacked star image ($F_{S,n,m}$) at the corresponding object's rate of
motion on a night-by-night basis to create an object-specific
synthetic asteroid PSF.

To create the trailed PSFs we created $2N+1$ shifted sub-images
($F_{S,n,m,k}$ with $-N \le k \le N$) of the stacked stars.  The shift
in pixels for each sub-image $k$ is given by:
\begin{eqnarray*}
\Delta x_k &=& \frac{k}{2N}\:\frac{1}{s}\:\frac{\Delta \alpha}{\Delta t} \\
\Delta y_k &=& \frac{k}{2N}\:\frac{1}{s}\:\frac{\Delta \delta}{\Delta t} \; {\rm ,}
\end{eqnarray*}
\noindent where the pixel scale $s = 0.185 ^{\prime\prime}$/pixel and we used
$N = 5$ (\ie 11 sub-images).  We need not consider cross terms because
the CFHT MegaCam $(x,y)$ axes are precisely aligned with
(RA,Dec)=($\alpha,\delta$) but we did take into account the
$\cos(\delta)$ term for the motion in RA.  The flux in each shifted
stellar profile thumbnail ($F_{S,n,m,k}$) was then combined such that
the flux in pixel ($i,j$) in the trailed, unnormalized asteroid model
thumbnail ($F'_{A,n,m}$) for object $n$ in image $m$ is
\begin{displaymath}
F'_{A,n,m}(i,j) = \frac{1}{2N+1}\: \Sigma_k \:F_{S,n,m,k}(i,j) \:.
\end{displaymath}
\noindent $F'_{A,n,m}$ was then normalized by its total flux within a
$2.0^{\prime\prime} \;$radius of the center to create the synthetic asteroid
model ($F_{A,n,m}$) specific to each TALCS object and image.  The
models were then median-combined to sub-pixel accuracy across the
images with seeing $< 0.8^{\prime\prime} \;$and normalized to create a
synthetic asteroid model specific to each TALCS object, $F_{A,n}$.

\subsubsection{Constructing the Coma Model: $F_{C,n}(i,j)$}

We also constructed one synthetic coma model ($F_{C,n}$) per TALCS
object $n$ using a method similar to the construction of the asteroid
model (see \S 4.1.2).  We convolved a spherically symmetric $r^{-1}$
coma profile with each synthetic asteroid model before stacking
($F_{A,n,m}$) to create an unnormalized, image- and object- specific
coma model $F'_{C,n,m}$.  These models were then normalized by the
total flux within the central $2.0^{\prime\prime} \;$ radius, median-combined
across images as for the stacked TALCS objects themselves, and then
once again normalized to produce the final object-specific coma model,
$F_{C,n}$.

\subsubsection{Fitting the Models}

Fitting the
stacked target object image to the linear combination of the stacked
asteroid and coma models in eq. \ref{eq.FnFaFc} requires an error
model for the image(s).  The photometric error ($E_{n,m}(i,j)$) on
each pixel of $F_{n,m}$ --- the normalized thumbnail for each TALCS
detection before median combining --- is the square root of the raw
photon count including the background, normalized by the total flux
from the background-subtracted image.  We then median-combined
$E_{n,m}(i,j)$ using only those images $m$ for which the seeing was
$\leq 0.8 ^{\prime\prime}$ (as with the construction of the stacked target
images and their asteroid and coma models) to produce
$\tilde{E}_{n}(i,j)$.  The error on each pixel of the stacked target
object image is then:
\begin{displaymath}
E_{n}(i,j) = \frac{1.253}{N (i,j)}\:\: \tilde{E}_{n}(i,j) \:\: ,
\end{displaymath}
\noindent where a standard factor of 1.253 is included to account for
combining the median rather than the mean of the images and $N(i,j)$
is an integer array of the number of images included in the stack at
each pixel.  $N(i,j)$ is pixel-dependent because centroid shifting in
the stack causes the thumbnails from different images to not overlap
perfectly.  A parallel method was used to determine errors for the
synthetic asteroid and coma model images, $E_{A,n}(i,j)$ and
$E_{C,n}(i,j)$, respectively.

The fitting algorithm assumes that the model, the right hand side of the equation,
 is error-free.  We incorporated all the error into
our stacked target object image as
\begin{displaymath}
E'_{n}(i,j) = \sqrt{ E_{n}^2(i,j) + E_{A,n}^2(i,j) } \: .
\end{displaymath}
\noindent The coma model's contribution to the error, $E_{C,n}(i,j)$,
is negligible because it is a convolution of a
perfectly symmetric, error-free model.

\subsubsection{Dealing with Systematics}
\label{ss.DealingWithSystematics}

Despite the care devoted to creating customized PSF models for each
object we found that our resulting formal errors on the fits to the
stacked asteroid thumbnails could not compensate for all the
systematic problems to the technique.  \ie the formal errors
on the derived $f_c$ values suggested that a large fraction of the
objects had significant coma and mass loss rate despite there being no
obvious visually detectable coma.  Instead, like the tail detection
method described above, we resorted to using a ranking method.
Ranking statistics are more robust than parametric statistics because
they do not assume any properties of the data (\eg Gaussian PSFs).
Ranking each TALCS object's coma fraction requires a set of similar
objects that have no coma.  We constructed the coma-free asteroid
comparisons in the manner described below.

First, we defined a metric $Z$ to quantify the similarity between 1) a
target stacked asteroid image with magnitude $\tilde{m}$ and rates of
motion in the $x$ and $y$ direction of $\tilde{\dot x}$ and
$\tilde{\dot y}$ respectively and 2) detections of other asteroids ($k$)
with magnitude $m_{k,i}$ and rates of motion ${\dot x}_{k,i}$ and
${\dot y}_{k,i}$ in image $i$:
\begin{displaymath}
Z_{k,i}^{2}\:= \: \frac{(\;      m_{k,i}  - \tilde{m}     \;)^{2}}{a^{2}} \: 
           + \: \frac{(\;{\dot x}_{k,i} - \tilde{\dot x}\;)^{2}}{(b/2)^{2}} \: 
           + \: \frac{(\;{\dot y}_{k,i} - \tilde{\dot y}\;)^{2}}{(b/2)^{2}} \:\:,
\end{displaymath}
\noindent The values of $a$ and $b$ were determined empirically to be
the inverse of the limit at which PSFs of different magnitudes and
trailing rates were similar enough to combine ($a=0.2$~differential
magnitudes and $b=0.25$~pixels).  

We then compiled a list of the $N^{2}$ smallest $Z$ values for each
stacked target asteroid where $N$ is the number of frames that were
used to construct the stacked target image and randomly selected $N$
frames from these closest matches such that no more than 20\% of the
frames came from the same TALCS object and there were no duplicates.
Combining the randomly-selected comparison frames renders a stacked
image comparable to that of the target and absent of any coma unless
more than 20\% of the main belt asteroids with $H < 21$ display comae (which is
highly unlikely given the rate of discovery for these sizes; \citealt{hsi06b,gil09}).
We call each of the generated stacked no-coma images a `null-image'
and could create as many null-images as necessary for each stacked
target asteroid.

Then, to quantify the significance of the derived $f_c$ for each
stacked target asteroid we computed $h$, the percentile under which
the target's $f_c$ falls compared to the null-images (see \S5.2 for a 
detailed discussion of $h$).  In the absence
of coma the distribution of $h$-values for a stacked asteroid image
should be uniform.

\subsection{Sensitivity}
\label{ss.ComaSearchSensitivity}

To determine the mass loss rate sensitivity of our coma detection
method we generated 10,000 null-images each at magnitudes
between 17 and 23 and at three heliocentric distances of
$R=1.6,2.6,3.6$~AU with corresponding geocentric distances of
$\Delta=R-1$.  For each set of null-images we used the coma fitting
procedure described above to determine $f_c$ and used the $10^{\rm
  th}$ largest $f_c$ from each set ($f_{c,limit}$) to represent the
$p=0.001$ statistical significance level - \ie there is a 1 in 1000
chance that a random stacking of assorted randomly chosen asteroids with mutually 
similar magnitudes and trailing rates) would produce $f_c > f_{c,limit}$.  
Then we converted $f_{c,limit}$ to
a flux (or coma apparent magnitude) from which we derived a mass loss
rate using Equation \ref{eq:eq.dMdt}.  Figure
\ref{fig:comaSensitivity} shows the $f_c$ $p=0.001$ coma fraction
(top) and the corresponding mass loss sensitivity limit (bottom) as a
function of magnitude for the three values of $R$.  For $R=2.6$~AU,
our sensitivity limit is typically better than 0.1 $\rm kg\, s^{-1}$,
although it varies by over an order of magnitude between $R=3.6$~AU
and 2.6~AU.  Thus, our coma identification technique is sensitive to
mass loss rates comparable to the known MBCs listed in Table
\ref{tab:knownmbcs}.

\section{Results \& Discussion}
\label{s.ResultsAndDiscussion}

\subsection{Tail Search}

\begin{figure}
\includegraphics[height=1.65in,angle=270]{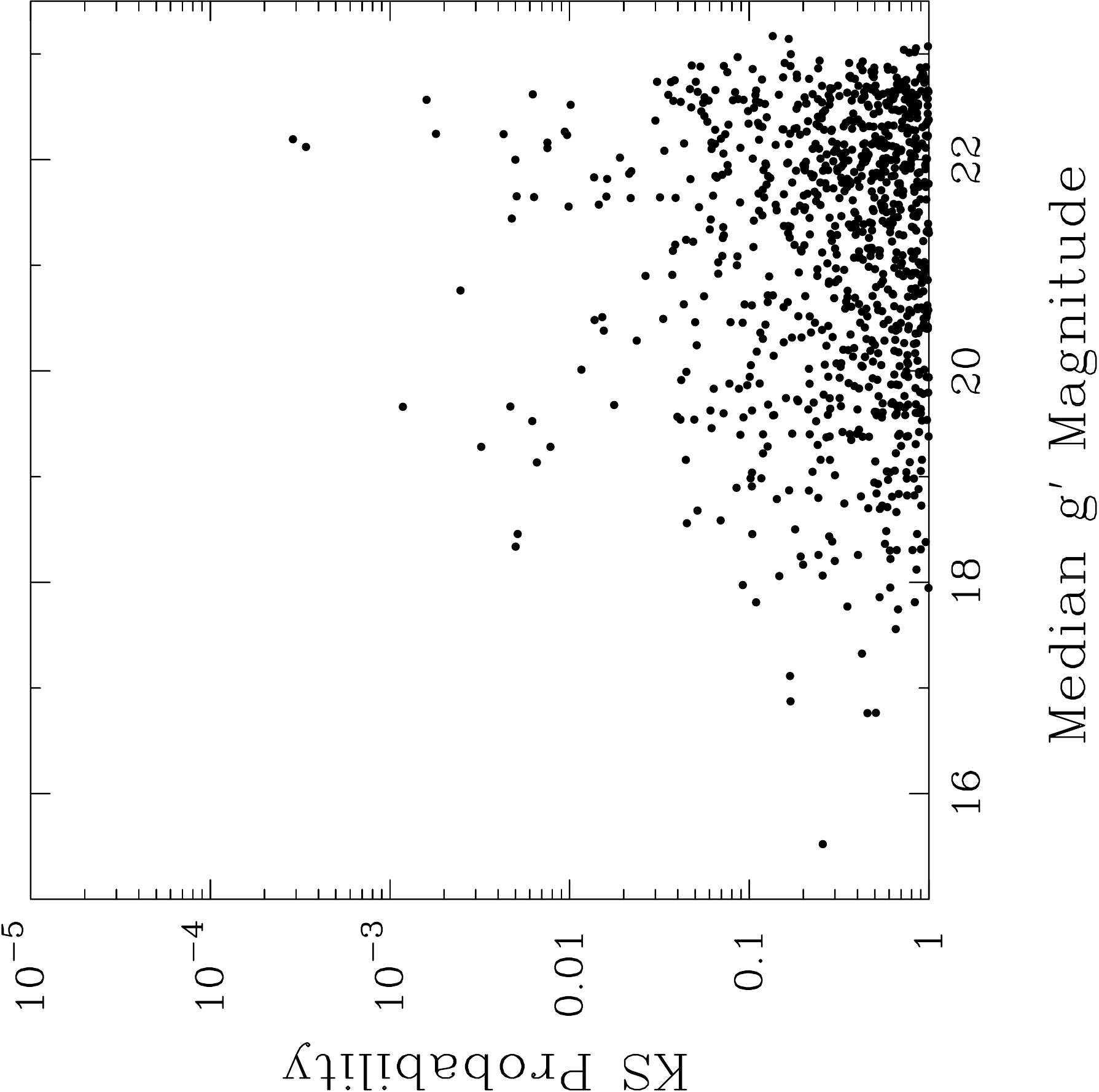} \quad
\includegraphics[height=1.65in,angle=270]{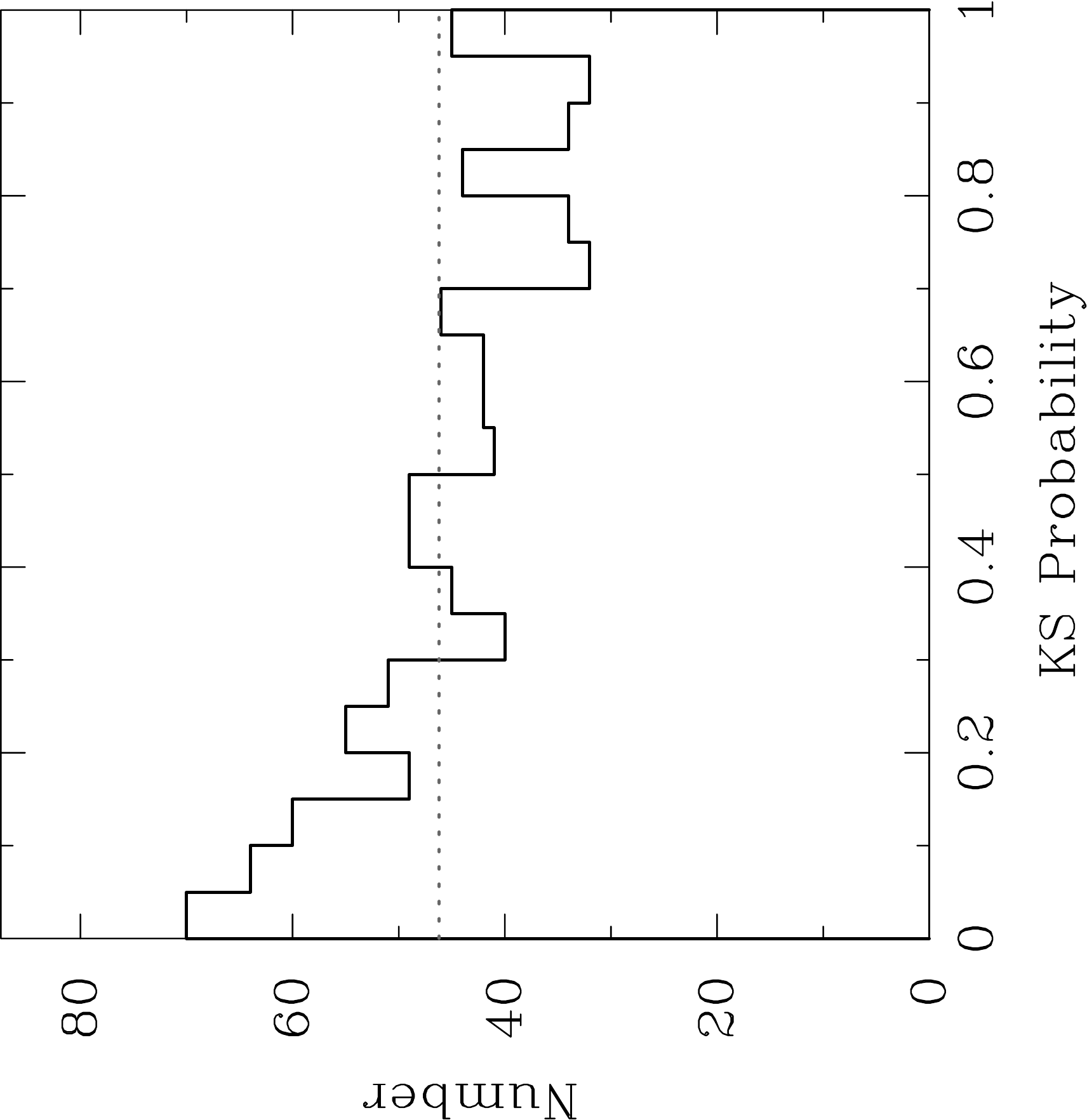} \quad
\includegraphics[height=1.65in,angle=270]{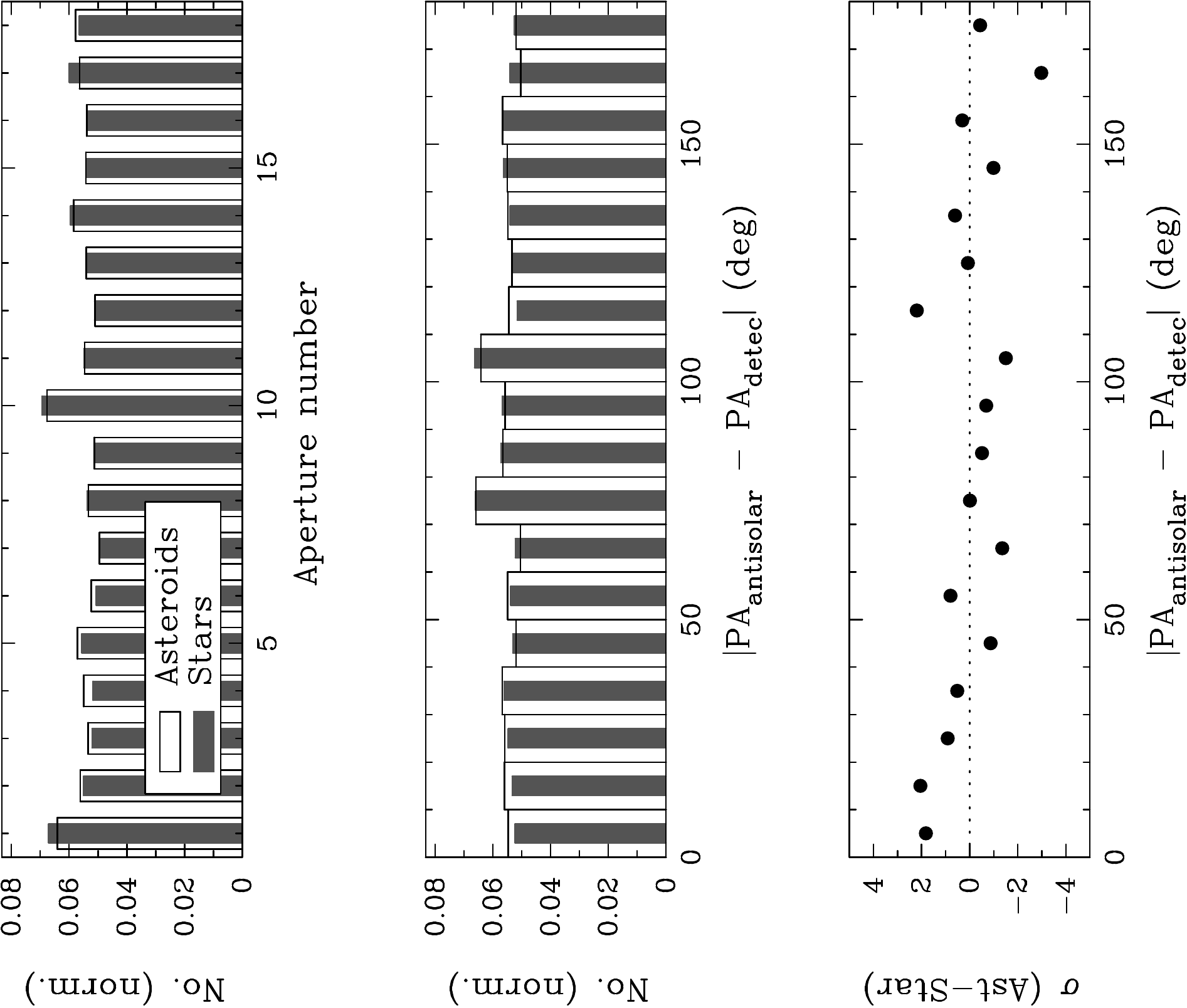} 
\caption{\label{fig:talcssliceresult} Left: Tail detection
  significance for all asteroids in the TALCS data set.  The vertical
  axis is the Kolmogorov-Smirnov (KS) probability as in Figure
  \ref{fig:slicemethodstrength}.  Center: The distribution of KS
  probabilities from the left panel.  The horizontal dotted line
  represents the expected distribution under the assumption that there
  is no tail or trail activity. Right: The angular distribution of the
  brightest detection segment as a function the of angular aperture
  number (top) and deviation from the expected anti-solar tail
  direction (center) for asteroids and similar stars ($\pm 0.1$ mag)
  on the same chip; and (bottom) the significance in $\sigma$ of the
  excess asteroid counts in each angular bin compared to the stars.
  It is evident from the stars that the angular distribution is
  non--uniform and dominated by systematics (top and center), but
  there is $\sim 2\sigma$ excess in asteroids for the two bins closest
  to the antisolar direction once the systematics are removed using
  the stars. }
\end{figure}

Figure \ref{fig:talcssliceresult} shows the results of applying our
segmented annulus tail detection technique described in
\S\ref{s.SearchForCometaryTails} to all the asteroids in the TALCS
data set. The strongest detection is at the KS probability level
$p=2.9\times10^{-4}$ which is expected to randomly occur about one
third of the time in a sample of nearly 1000 asteroids.  We conclude
that we find no evidence of MBC tail activity in any individual
asteroid within the TALCS data set.  This result may not be surprising
because half the TALCS sample has $H_v>17.7$ and thus are smaller than
the smallest known MBC.  Smaller objects may be less likely to have
tails because there is less volume to store the volatiles and less
regolith to protect buried volatiles from seasonal thermal effects.

It is important to note that the center panel of Figure
\ref{fig:talcssliceresult} shows that the distribution of KS
probabilities is biased to low probability events.  But a numerical
simulation of 1000 stars with uniform randomly chosen brightest-slice
rankings $f$ shows that the distribution should be uniform.  Applying
a KS test to the tail detection KS probability distribution shows that
they are inconsistent with a uniform no-activity null hypothesis at
the $p=1.2\times10^{-5}$ level, unlike the 1000 stars.

The right panel of Figure\ref{fig:talcssliceresult} shows the angular
distribution of the brightest detection segment compared to the
distribution for matched stars. Although the distributions for both
stars and asteroids are non--uniform, possibly from angular
asymmetries in the telescope optics, the bottom panel of the right
panel shows that there is a $\sim 2\sigma$ excess of asteroids over
stars in the $|{\rm PA}_{\rm antisolar}-{\rm PA}_{\rm detec}|\approx
0^\circ$ direction.  \ie the brightest segment is aligned with the
expected antisolar tail direction.  If the sample is restricted to
asteroids where the total KS probability is limited to $p<0.05$ or
$0.01$, the excess at zero angle remains at the $1.5$ to $2.0\sigma$
level.

Thus, barring any systematic biases that we have not already taken into
account, we conclude that there is evidence for low level excess directionalized
flux around many of the TALCS asteroids and, generalizing further,
that main belt asteroids as a class exhibit weak tails or trails.
Unfortunately, the evidence is too weak to identify specific asteroids
in our sample because none have $p \ll 0.001$.  In a sample of 1000
asteroids we expect one object to have $p = 0.001$ by chance so the
small KS objects in our TALCS data are not likely candidates.

While we can not select individual objects as candidates for faint MBC
tails we can estimate that there is an excess of about 50 objects with
faint tails in the TALCS data set.  In other words about 5\% of the
main belt objects in the TALCS data set might exhibit very faint tail or
trail activity.  If the result is correct it implies that deep images
of a large sample of main belt objects acquired in a manner suitable
for analysis with our technique would identify low-level activity in
main belt objects that otherwise appear asteroidal.

Considering that the known MBCs are transient in their activity our
result that $\sim$5\% of asteroids are active at {\it any} time
implies that a large fraction of main belt asteroids could be active
at {\it some} time.  This result further blurs the traditional
distinction between asteroids and comets.  On the other hand, we
speculate that the mechanism driving the low-level tail activity might
not be volatile driven ejection but the result of regolith ejection
due to impacts of small meteoroids.  In this case, our results offer
constraints on the main belt collisional environment such as the amount of 
primordial implantation of ice-rich bodies and the fraction of 
those bodies that were collisionally disrupted \citep[\eg][]{bot05,obr05}.

\subsection{Coma Search}

\begin{figure}
\includegraphics[width=5.5in, height=5.5in]{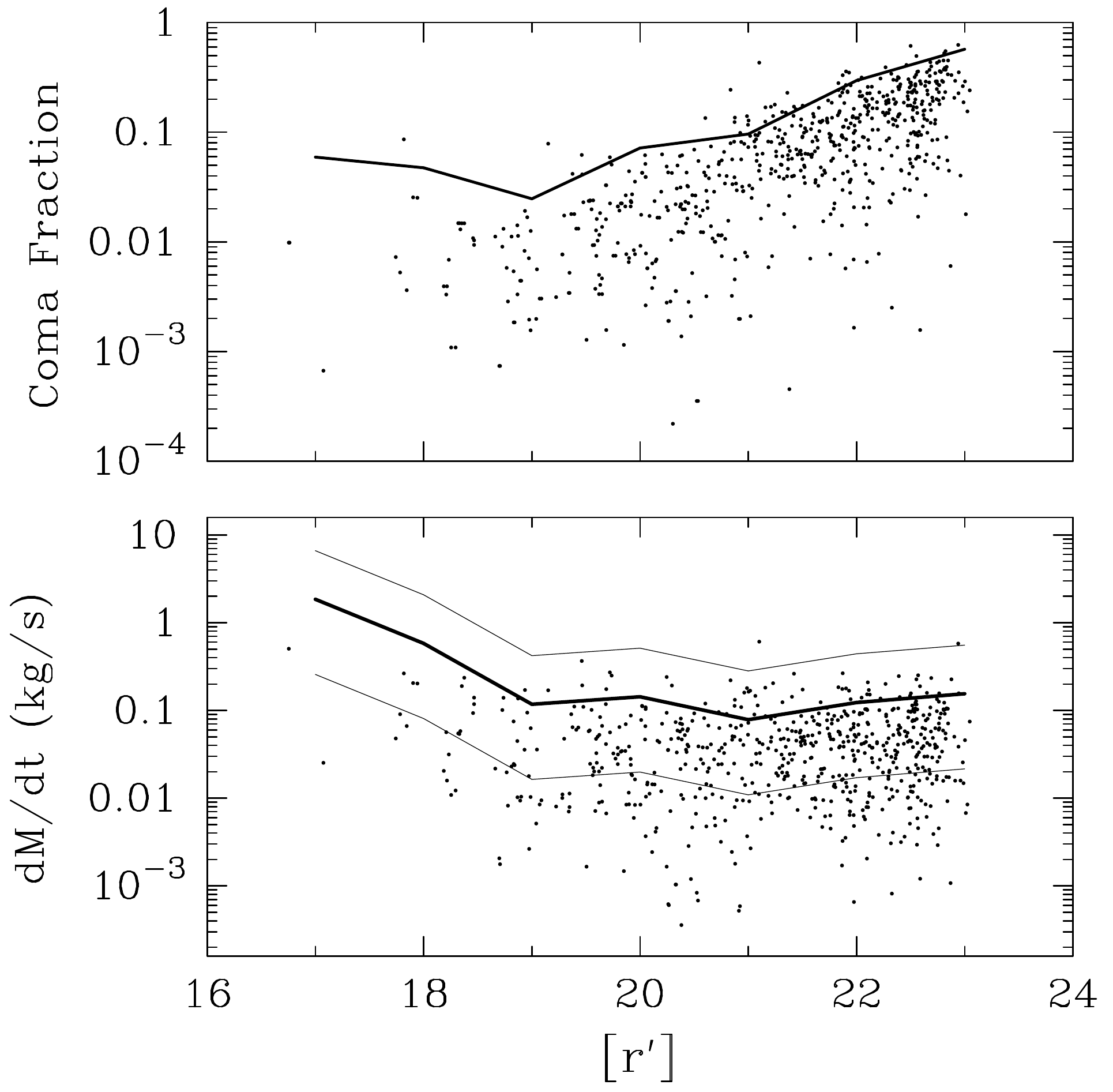}
\caption{\label{fig:comaSensitivity} Top: The solid line is the 99.9\%
  confidence bound on the coma fraction $f_c$ for a benchmark asteroid
  of a given magnitude.  Randomly stacking asteroids yields an $f_c$
  above the solid line only 0.1\% of the time.  The data points
  represent the derived values for the TALCS asteroids.  Bottom:
  A conversion of the $f_c$ in the top panel into a mass loss rate.
  From top to bottom the solid lines represent placing the benchmark
  asteroid at heliocentric distances of $R=3.6,2.6,1.6$~AU.}
\end{figure}

We applied the technique of \S\ref{s.SearchForCometaryComae} to fit
each of the TALCS objects to an asteroid and coma model and thereby
measure the fractional contribution of the putative coma ($f_c$) to
the objects's flux.  Figure \ref{fig:comaSensitivity} shows that most
of the resulting $f_c$ were below our sensitivity threshold and
effectively consistent with zero.  But that sensitivity threshold is
sensitive to object-specific parameters such as magnitude, rates of
motion, PSFs, and the number of frames used in creating the stacked
asteroid.

Thus, for each stacked asteroid the null-image stacking procedure
described in \S\ref{ss.DealingWithSystematics} was repeated 50 times
per target.  In the absence of coma the distribution of $h$-values for
all TALCS objects should be uniform so that after 50 trials no more
than $1/50^{th}$ of the TALCS objects should have $h=1.0$, as discussed in \S4.1.5.  But
instead of the $\sim 20$ expected objects we found 34.  We then
subjected those 34 objects to an additional 500 trials after which we
expected only $\sim 2$ TALCS objects with $h=1.0$ but 8 objects fell
into this bin.  An additional 5000 trials for each of those 8 objects
should have produced zero objects with $h=1.0$ but 3 remained.  In
other words, at this stage it appeared that we had identified 3 MBC
candidates --- the problem was that under detailed visual examination
none of these objects displayed any visible coma.

To further investigate the 8 candidates with very high $h$-values
($h>0.998$) we divided the set of images for each object in half and
separately processed each half with the detection and sensitivity
algorithms.  If the positive detections remained in each half of the
data it would suggest the coma is real.  On the other hand, if the
coma candidates were not detected in both halves of the data it would
suggest either systematic errors in the detection and/or sensitivity
algorithms or a transient phenomenon resembling a coma.  

Only 1 of the 8 objects had consistently positive coma detections in
both halves of its data and, upon visual inspection, it was found to
exhibit transient coma-like phenomena in 2 of its 6 images, one in
each half.  The transient phenomena may be due to it passing over a
faint background source.  Since we expect real comae to persist over
the two weeks spanned by the TALCS observations, and since the
signal-to-noise was poor in both the individual and combined frames
for this object, we conclude that it is not a MBC candidate.  Thus, we
have a null result in our search for cometary-like comae in the TALCS
survey.

The derived mass loss rates for the TALCS objects are shown in
Fig. \ref{fig:comaSensitivity} and it is no surprise that our derived
mass loss rates are small and below our sensitivity limits.  Typical
comets have mass loss rates in the range $10^{-3} \lesssim \dot{M}
\lesssim 10^3$ kg$\cdot$s$^{-1}$ \citep[\eg][]{lam04} with the MBCs
falling in the smaller decades of the distribution but still within
the range of our sensitivity limit of about $10^{-1}$~kg$\cdot$s$^{-1}$.

\subsection{Upper Limits on MBC Orbit and Size Distributions}

\citet{hsi06a} estimated that there is one active MBC for every 300
main belt asteroids based on their Hawaii Trails Survey that focused
primarily on Themis family members in the outer main belt.  That
family and heliocentric range were of particular interest because the
MBCs EP and 176P were already known to be Themis family members and it
seemed reasonable to expect that if sub-surface volatiles could
survive since the formation of the solar system they would most likely
do so in objects that are farther from the Sun.  Considering that
\citet{hsi06a} did not account for these observational selection
effects the 300:1 ratio of main belt asteroids to active MBCs is a
lower limit --- the ratio could be considerably larger when averaged
over the main belt as suggested by \citet{gil10}.  However, at face
value, the estimate suggests that we should identify $\sim 3$ MBCs
within the TALCS sample since our detection techniques are sensitive
to the same cometary activity levels as the known MBCs.

Our null result allows us to set new upper limits on the number
distribution of MBCs in absolute magnitude, semimajor axis,
eccentricity, and inclination by employing the technique of
\citet{mos08}.  Given the false-positive rate ($F$), the differential
absolute magnitude distribution of TALCS objects, $n(H)$, the
probability of detecting an active MBC within the survey ($P_d$), and
the completeness of the survey as a function of absolute magnitude,
$C(H)$, the actual number of MBCs as a function of absolute magnitude
is given by:
\begin{equation}
N(H) = \frac{(1-F) \;\; n(H)}{P_d \;\; C(H)} \;\;.
\end{equation}
\noindent We take $F = 0.001$ ($\sim$1/924) because of the one
questionable detection described at the end of
\S\ref{ss.ComaSearchSensitivity} though it is functionally equivalent
to a zero false-positive rate.  Assuming activity levels similar to those 
previously observed in current MBCs, $P_d = 1.0$ because we have
demonstrated that our technique is capable of detecting MBC tails
and/or comae at all absolute magnitudes within our sample.

We used the ASTORB database to obtain the true number distributions
($N(x)$ with $x = a,e,i,H$) of all main belt objects to an absolute
magnitude of $H < 14.8$ with $e < 0.4$, $i < 45.0^{\circ}$, and 2.0~AU~$<
a < 3.5$~AU.  This sample of known asteroids is believed to be
complete (\citealt{bow07}), \ie all main belt asteroids with $H<14.8$ are
thought to be known.  We extrapolated to $H=21$, the limit of the
TALCS sample, using \citet{jed02}:
\begin{equation}
N(H) = 0.0059\times 10^{0.5*H} ,\;\;\;\;\;\; 14.8 \le H < 21.
\end{equation}
We then derived $C(x)$ by dividing the \textit{observed} number of
objects within a given $H$ bin by the total number of objects in that
bin.

The unbiased differential number distribution in orbit element $x$ ($x
= a,e,i$) is then given by
\begin{displaymath}		    
N(x) = \frac{A(H)\;\; (1-F) \;\; n(x)}{P_d \;\; C(x; H < 21)} \;\;, 
\end{displaymath}
\noindent where $n(x)$ is the observed number distribution for our
TALCS objects and $A(H)$ is a normalization factor computed from the
$n(H)$ cumulative number distribution so that the integrated number of
TALCS objects in our range of $x$ equals the number in our range of
absolute magnitudes.  \ie $A(H) = \int n(x) dx / \int n(H) dH$.
$C(x; H < 21)$ is the completeness of the survey as a function of
orbit element $x$ for $H < 21$.

\begin{figure}
\begin{center}
\includegraphics[width=5in, angle=0]{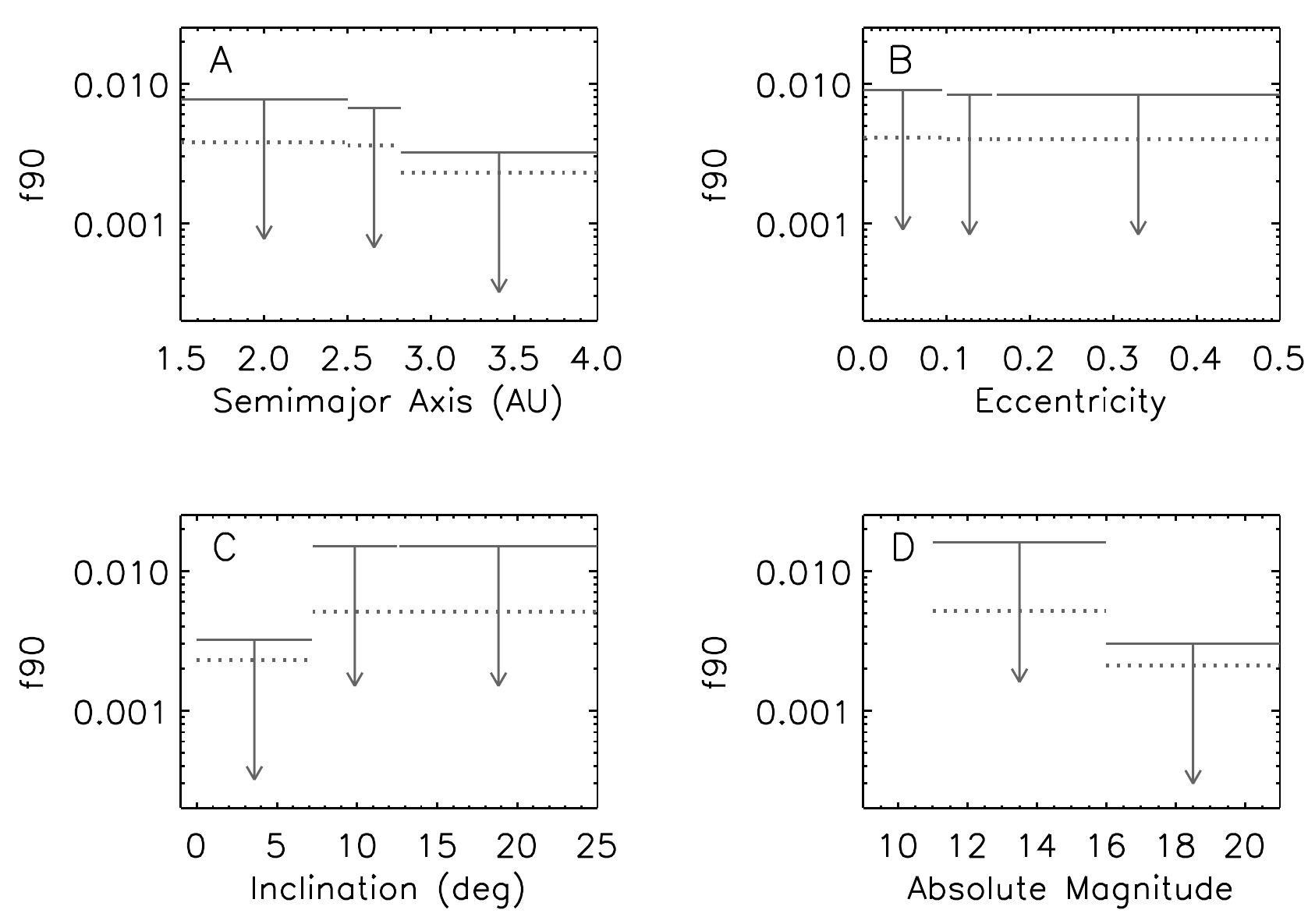}
\caption{Poisson (solid) and Bayesian (dotted) 90\% confidence limits
  on MBC number distributions as a function of (a) semi-major axis,
  (b) eccentricity, (c) inclination, and (d) absolute magnitude.  See
  \ref{appendix:stats} for a detailed discussion of the
  Poisson vs. Bayesian approach.}
\label{fig:upperlimitdist}
\end{center}
\end{figure}

Since we found no MBC candidates, $n(x)=0$ for all $x$, and
Fig.~\ref{fig:upperlimitdist} gives the 90\% confidence bound limit on
the fraction of MBCs using both a standard Poisson framework and a
Bayesian framework that accounts for our prior belief of a 1/300
fraction (see \ref{appendix:stats}). The latter method
effectively estimates the MBC fraction by transforming \citet{hsi06a}'s 1/300 to
1/(300+924).

We found that the MBC:MBA ratios for $H<21.0$ are $\lesssim$1:300,
$\lesssim$1:350, and $\lesssim$1:500, for the inner, middle, and outer belt
respectively and, averaged over the entire main belt,
MBC:MBA$\lesssim$1:400.  While these limits are numerically not markedly
different from (\citealt{hsi06a}) they are applicable to asteroids to much
smaller sizes ($H<21$ vs. $H<18$) and were derived from a survey in
which we explicitly accounted for observational selection effects.
Indeed, \citet{hsi09b} recognized the bias in the earlier result and
revised the applicability of the MBC:MBA ratio reported in \citet{hsi06a} 
to low-inclination ($i < 3.0$), `km-scale' MBCs in the outer belt
--- a population composing only $\sim 10$\% of the TALCS sample.  
\citet{hsi09b} further re-evaluated the MBC:MBA ratio based on the Hawaii 
Trails Project (HTP) to be $\sim 1:100$.

\citet{gil09} and \citet{gil10} used serendipitous detections of asteroids in the CFHT
Legacy Survey \citep[CFHTLS,][]{jon06} to identify MBC candidates and
measure the MBC:MBA ratio.  Since the CFHTLS was not explicitly
targeted at identifying MBCs they could correct for the survey's
observational selection effects.  They identified one MBC candidate after
examining $\sim25000$ main belt objects and thus concluded that the
MBC:MBA ratio is $\sim$1:25000 for objects with $H < 16$
(corresponding to objects of more than a few kilometers in diameter).
While their value is considerably more stringent than ours it applies
to objects $\sim$10 times the diameter of our study.  It will be interesting to eventually 
measure the fraction of MBCs as a function of their diameter (or $H$) in an
effort to provide constraints on thermal models for asteroid evolution
but, in order to do so, there will need to be a larger sample study
similar to this work or that of \citet{gil09} and \citet{gil10}.

\subsection{Constraints on the Snow Line}

One of our original motivations for the TALCS MBC survey was to
identify a handful of MBCs in a relatively unbiased survey --- at
least relatively unbiased in comparison to the HTP of \citet{hsi06a}
that specifically targeted regions of the main
belt where they thought it likely to discover MBCs.  Based on their
1:300 MBC:MBA ratio we optimistically expected to identify a few
 MBCs in the TALCS data set.  With such a set of MBCs in hand we
thought it would be possible for the first time to constrain
the location of the primordial snow line. \ie if we found 5 MBCs
beyond 2.5~AU and none in the inner belt it could have been strong
evidence for a snow line around 2.5~AU.  If radial mixing is negligible, then 
the positions of MBCs could give hints as to the primordial snow line's 
location.  Having identified zero MBCs
in this study we are limited in its application to the location of the
snow line but, under a set of simplified assumptions, it is possible
to use the locations of the known MBCs to estimate its position.

If the snow line is at semi--major axis $a_s$ and one assumes that 1)
{\it observable} asteroids are evenly distributed in semi--major axis
$a\in[A_0,A_1]$ in the asteroid belt (this is in fact the case as
shown in fig. \ref{fig:dist_TALCS_vs_AST} for $A_0\sim2.2$ and
$A_1\sim3.2$), and 2) $n$ MBCs have been observed {\it in an unbiased
  manner} at semi--major axes $a_i$, (the unbiased assumption was the
motivation for using the TALCS survey), then the probability that $a_{\rm min}$,
 the smallest value of $a_i$, is greater than some value of
$a$ is given by the binomial probability:
\begin{equation}
P( a_{\rm min} >a | a_s) = 
\left\{
\begin{array} {c@{\quad}l}
{\displaystyle 
   \left({A_1  -  a \over A_1 - \max(a_s,A_0) } \right)^n } & \hbox{for\  }  \max(a_s,A_0)<a<A_1  \\
    0 & \hbox{for\  } a\ge A_1 \\
     1 & \hbox{for\  } a\le  \max(a_s,A_0) 
\end{array}
\right.
\end{equation}
Then the differential probability distribution of  $a_{\rm min}$ is given by
\begin{equation}
  P(a_{\rm min} | a_s) 
    =  - {d  P( a_{\rm min} >a | a_s) \over da}  {\Big|}_{a=a_{\rm min}}
    =  {n \left(A_1  -  a_{\rm min}\right)^{n-1} \over \left[ A_1 - \max(a_s,A_0)  \right]^n }.
\end{equation}

With no prior expectation for the position of the snow line we simply
assume that the probability is constant for finding $a_s$ over some
range of $a$ that encompasses the asteroid belt.  From the Bayesian
perspective, $P(a_{\rm min} | a_s) \propto P(a_s | a_{\rm min})$, where
$a_{\rm min}$ is now the innermost observed MBC.  Substituting
$A_1=3.2$~AU and using $a_{\rm min}=2.72$~AU from Table
\ref{tab:knownmbcs} as the innermost reliable MBC we obtain a
probability for the location of the snow line
\begin{equation}
P( a_s | a_{\rm min}=2.72 ) = 
\left\{
\begin{array} {c@{\quad}l}
{\displaystyle 
  {\rm const} \times  \left[ 3.2 - \max(a_s,2.2)  \right]^{-6}} & \hbox{for\  } a_s<2.72\\
  0 &  \hbox{otherwise}
\end{array}
\right.
\end{equation}
where the distribution is to be normalized over the plausible range of
$a_s$.

For example, if we believe that any $a_s>1.5$~AU is equally plausible
then from the existing 6 MBCs we derive a nominal 85\% confidence that
the snow line is outside $2.5$~AU.  Although this is a weak result
given the current ensemble of MBCs it provides a formal mechanism to
gauge how the estimate will improve as the MBC sample grows.  It is also
consistent with conclusions from theoretical models and meteorite studies that
in the first few million years of the solar system's lifetime, the snowline was 
located inside of $\sim 3$~AU \citep[\eg][and references therein]{ber11}.

Our naive assumption that the location of {\it the} snowline was
frozen in place during the solar system's formation and might still be
present in its primordial location may be of limited utility.
\citet{gar07} and \citet{min11} calculated that during the formation of the inner
solar system the snow line migrated from as little as 0.7~AU to many
tens of AU due to changing dust opacities and mass accretion rates in
the evolving primordial nebula.  Based on their work we might expect a
gradient of subsurface water ice abundances across parent body sizes,
local number densities, semi-major axes, etc. and, hence, varying MBC
activity levels throughout the main belt.  In any event, given the
lack of knowledge about the evolution of the snowline in proto-solar
nebulae, attempts to identify its location or to measure the water
gradient in the solar system could provide useful constraints on
modelling solar system formation.  Perhaps the solar system maintains
a record of evolutionary transitions in the proto-solar nebula as the
snowline advanced further from the Sun much like glacial moraines
leaving permanent records of changes in climate as glaciers recede.

The ability to constrain the location of the primordial snow line in
our own solar system is further complicated by recent revelations
involving massive reorganization of the distribution of the small
bodies during early solar system evolution.  The ability to constrain
the location of the snow line presupposes that the current
distribution of small bodies is mostly representative of their
primordial distribution.  The observed taxonomic distribution
of small bodies progressing from S-complex asteroids in the inner
belt, to C-complex in the middle and outer belt, to D- and P-type
asteroids in the Hilda and Jupiter Trojan regions may represent a 
relic of their original distributions.  

\citet{lev09} have thrown this assumption entirely upside down in
suggesting that an inward-then-outward migration of the giant planets
caused planetesimals from beyond Neptune's orbit to scatter into the
main belt region, particularly the outer belt.  In this scenario
objects currently in the main belt do not represent the primordial
distribution of planetesimals nor do nearby objects in semi-major axis
necessarily have a common origin and formation mechanism.  If their
work is correct, observations of cometary activity in the main belt
could be more useful for mapping out the redistribution of material
from the outer solar system than constraining the primordial or
contemporary snow line.

\section{Summary}

\begin{enumerate}
\item We developed quantitative techniques for sensitive searches for
  faint tails and coma around objects that otherwise appear
  asteroidal.  The tail detection algorithm identifies excess flux in
  pie-sgements of an annulus around an object.  The coma detection
  algorithm fit a detailed PSF model to each candidate object that
  included a point-like and coma-like PSF component.  The tail
  detection technique could easily identify MBCs at the same activity
  level as the known MBCs.  The coma detection technique is sensitive
  to mass loss rates that are comparable to the known MBCs as well.
  Both techniques can be applied to large scale sky survey detections
  of asteroids.
\item We did not identify any MBC candidates using either technique.
\item Using our null-detection result we set upper limits on the
  number of MBCs as a function of semimajor axis (AU), eccentricity,
  inclination, and absolute magnitude.
\item We determine that the MBC:MBA ratio for the entire belt to
  $H<21.0$ ($\sim$150~m in diameter) is $\lesssim$1:400.  For the
  inner, middle, and outer belts, the ratios at $H<21.0$ are no
  greater than 1:300, 1:350, and 1:500, respectively.
\item We presented evidence for extremely low-level tail activity in a
  surprising fraction of apparently generic main belt asteroids ---
  about 5\% of our asteroid sample showed evidence of directional
  excess flux suggesting that many `asteroids' display low-level
  activity.  If correct, it is not clear whether the activity is due
  to small scale impacts liberating regolith or to sub-surface
  volatile activation.  It might be possible to identify this activity
  in targeted deep imaging of many main belt asteroids.  It might also
  be possible to study the production mechanism by breaking a large
  sample of objects into sets of objects near aphelion and perihelion.
  If the excess activity is related to the objects's mean anomaly it
  could indicate thermal-induced volatile activity.
\item We present a Bayesian argument based on the known distribution
  of MBCs that suggests with 85\% confidence that the contemporary
  snow line is beyond 2.5~AU.
\end{enumerate}

\vspace{1.0cm}

{\bf ACKNOWLEDGEMENTS}

This work was made possible with NASA PAST grant NNG06GI46G and was
based on observations obtained with MegaPrime/MegaCam, a joint project
of CFHT and CEA/DAPNIA, at the Canada-France-Hawaii Telescope (CFHT)
which is operated by the National Research Council (NRC) of Canada,
the Institut National des Science de l'Univers of the Centre National
de la Recherche Scientifique (CNRS) of France, and the University of
Hawaii.  This work was partly supported by NASA through the NASA
Astrobiology Institute under Cooperative Agreement NNA09DA77A issued
through the Office of Space Science.  We thank the people of Hawaii
for use of their sacred space atop Mauna Kea.

\clearpage

\begin{table}[!h,!t,!b]
\begin{center}
{\LARGE\caption{Parameters of the seven known MBCs: Object's name, semimajor
  axis ($a$), eccentricity ($e$), inclination ($i$), orbital period
  (P$_{orb}$), effective diameter (D$_{e}$), absolute V magnitude
  ($H_{v}$), and mass loss rate ($\dot{M}$).  Data comes from
  \citet{hsi04}, \citet{hsi06a}, \citet{jew09}, \citet{hsi09},
  \citet{jew10}, \citet{mor10}, and the Minor Planet
  Center (website: http://www.minorplanetcenter.net).  \label{tab:knownmbcs}}}
\begin{tabular}{l|ccccccc}
\hline\hline
Object & $a$ (AU) & $e$ & $i$ (deg) & P$_{orb}$ (yr) & D$_{e}$ (km) $^{\dag}$  & $H_{v}$ & $\dot{M}$ (kg s$^{-1}$) \\
\hline
133P/Elst-Pizarro & 3.16 & 0.162 & 1.39 & 5.62 & 3.8 & 15.9 & 0.01 \\
238P/Read & 3.17 & 0.253 & 1.27 & 5.63 & 0.6 & 17.7 & 0.02 \\
176P/Linear & 3.19 & 0.194 & 0.24 & 5.71 & 4.4 & 15.1 & {\bf . . .} \\
P/2008 R1/Garradd & 2.73 & 0.342 & 15.90 & 4.50 & 1.0 & $\gtrsim$ 17.8 & $<1.5$ \\
P/2010 A2/Linear$^{a}$ & 2.29 & 0.124 & 5.25 & 3.47 & 0.1 & 21.3 & 0.1-5 \\
P/2010 R2/La Sagra & 3.10 & 0.154 & 21.39 & 5.46 & 1.5 & 15.5 & {\bf . . .} \\
596 Scheila$^{a}$ & 2.93 & 0.165 & 14.66 & 5.01 & 113.3 & 8.90 & {\bf . . .} \\
\hline
\end{tabular}
\end{center}
{\small Note: $^{\dag}$ Using a geometric albedo of 0.134 characteristic of the outer belt
  (\citealt{jed98,kla92}) to calculate the diameter from the published
  absolute magnitudes.
  
  $^{a}$ Excess flux from these objects is most likely the result of asteroid collisions rather than recurrent activity typical in comets.}
\end{table}

\clearpage

\begin{table}
\begin{center}
{\LARGE\caption{Tail fluxes of three known MBCs as a fraction $f_T$ of the
  central point source and as a median $g^\prime$ magnitude in the
  brightest photometric segment, illustrated in Fig. \ref{fig:elstpizpie}.\label{tab:knownmbcstails}} }\vskip30pt
\begin{tabular}{l|ccc}
\hline\hline
Object       & $f_T$ & $g^\prime$ \\
\hline
Elst-Pizarro & 0.059 & 22.87 \\
P/Read       & 0.086 & 21.86 \\
176P         & 0.037 & 23.08 \\
\hline
\end{tabular}
\end{center}
\end{table}
\clearpage

\clearpage

\clearpage

\clearpage

\clearpage

\clearpage

\clearpage

\clearpage

\clearpage

\clearpage

\begin{appendix}

\section{Bayesian and frequentist statistics for zero
  detections \label{appendix:stats} }

In this work we present constraints on the true (unbiased) rate of
MBC activity using observations that yielded zero detections in $M$
observations.  Such an extrapolation is inherently problematic because
it depends strongly on assumptions of the incidence $f$ of MBCs or,
more formally, on the Bayesian prior $P(f)$.

\subsection{A frequentist approach}

The customary Poisson frequentist model begins with the fact that the
probability of observing $n$ MBCs given an expected number
$\left<n\right>$ is
\begin{equation}
P_{\rm Pois}(n)={\left<n\right>^n e^{-\left<n\right>} \over {n!}}
={(fCM)^n e^{-fCM} \over {n!}}
\end{equation}
where in the rightmost component we have defined the survey
completeness (sensitivity) as $C\in[0,1]$ and identified
$\left<n\right>=fCM$.  Under the frequentist paradigm, $f_{90}$, the
90\% upper confidence limit on $f$ is given by the implicit equation:
\begin{equation}
0.9 = {\int_0^{f_{90}} P_{\rm Pois}(n) \,df \over \int_0^1 P_{\rm Pois}(n) \,df}
\end{equation}
\noindent For $n=0$
\begin{equation}
f_{90} = -(CM)^{-1} \ln (1-0.9) 
\end{equation}

However, this model has several undesirable features.  First, it
assumes that when $n=0$ MBCs are found the expected number is
1\footnote{because for $n=0$, $P(f)=CM e^{-fCM}$; then
  $\left<n\right>=CM \left<f\right>$ and $\left<f\right>=\int_0^1 f
  P(f) \,df\approx (CM)^{-1}$.} irrespective of the size of the
sample.  \eg If we observe only 10 asteroids and our completeness is
$C=1$ then the analysis yields $f_{90}=0.23$ --- implying that we are
90\% confident that $<23$\% of asteroids are MBCs.  On the other hand,
the alternative interpretation is that we are 10\% confident that MBCs
represent more than 23\% of asteroids but this is easily seen to be
wrong because of prior knowledge from other surveys.

Next, the statistical implications are altered by binning the data.
If we observe $M=1000$ asteroids and $n=0$ MBCs we would compute
$f_{90}=0.0023$ for the entire sample, meaning that we have 10\%
confidence that a typical sample of asteroids contains 2.3 or more
MBCs.  However, if were to divide the sample into 100 semi--major axis
bins of 10 asteroids each and re--apply the statistics we would assign
$f_{90}=0.23$ to each bin.  \ie that each bin of 10 asteroids has a
10\% chance of containing more than 2.3 MBCs, an expected MBC count
that far exceeds what was obtained when the data were contained in a
single bin.

Both examples show the formal Poissonian $f_{90}$ does not represent
a genuine confidence limit (in the sense of betting odds) because we
overestimate the $10\%$ probability assigned to $f>f_{90}$.  In the
first case, this happens because we assign a prior probability that
ignores previous knowlege.  In the second case we pretend that each
bin is independent when we know that a failure to find an MBC in 99
bins means it is very unlikely to find one in the $100^{\rm th}$ bin.

A resolution to the problems inherent to the Poissonian frequentist
approach is a Bayesian approach to the question.

\subsection{A Bayesian approach}

A Bayesian approach remedies at least the first flaw described above
at the cost of `contaminating' our confidence intervals with knowledge
from other surveys. Indeed, the above Poisson approach was simply a
Bayesian method with a constant prior on $f$.

We begin our analysis by assuming a prior for $f$
\begin{equation}
\label{eq:bayesprior}
P(f)=
 \Bigg\{
 \begin{array}{ll}
  [-f \log(f_0)]^{-1} & \mbox{\qquad for \qquad} f\in[f_0,1] \\
  0                  &\mbox{\qquad elsewhere}
 \end{array}
\end{equation}
where $f_0\ll 1$ is the smallest allowed value of $f$ and is assumed
to approach $0$.  The $f^{-1}$ prior is the basis of Benford's law
(\citealt{benford38}) and implies that $f$ is equally likely to reside in
each decade of magnitude within the range of interest.  By allowing
$f_0 \rightarrow 0$ we assert an initial belief that MBCs are
extremely unlikely to exist.

Next, we modify our prior using the study of \citet{hsi06a} (hence
HJ06) who found one MBC in a targeted survey of $M_{HJ}=300$
asteroids.  Bayes' theorem informs us that an experimental result $E$
modifies our prior belief for $P(f)$ according to
\begin{equation}
\label{eq:bayestheorem}
P(f | E) = {P(E | f) \times P(f) \over P(E)}
\end{equation}

Here $E$ is an experiment consisting of an observation of some number
of MBCs $n$ in a given sample of $M$ asteroids.  Assuming completeness
or survey sensitivity $C\in[0,1]$ gives the binomial probability
distribution:
\begin{equation}
\label{eq:bayesexp}
P(E | f) = P(n | f; C, M) =  {M! \over n! (M-n)!} (Cf)^n (1-Cf)^{M-n}
\end{equation}
where we use the notation that items after a semicolon are fixed
parameters.  Then the probability of observing experimental result
$E$, or $n$ objects, is
\begin{equation}
\label{eq:bayespf}
P(E)=P(n;C,M) = \int_0^1 P(f)  P(n | f, C, M)\, df 
\end{equation}
Thus the posterior probability of $f$ given experiment $E$ is obtained
by combining Equations \ref{eq:bayesprior}, \ref{eq:bayestheorem},
\ref{eq:bayesexp} and \ref{eq:bayespf}
\begin{eqnarray}
\label{eq:bayespost}
P(f|E) & = & P(f|n;C,M) \\
       & = & {  (Cf)^n (1-Cf)^{M-n} \times f^{-1}  \over
            \int_{f_0}^1  (Cf)^n (1-Cf)^{M-n} 
            \times f^{-1} \,df  } \nonumber
\end{eqnarray}
where the normalization given by $-\log(f_0)$ and the factorial terms
have cancelled.  The denominator of equation \ref{eq:bayespost} is a
constant normalization term and for $n>0$ we may allow it to reach the
limiting value $f_0=0$ without encountering a singularity.

Using HJ06's result of $n=1$ the posterior probability is
\begin{equation}
\label{eq:hj06post}
P(f|E_{HJ})=M_{HJ} \times (1-f)^{M_{HJ}-1}.
\end{equation}
This probability is relatively constant for $f\lesssim 1/M_{HJ}$
compared to the original divergent prior $P(f)\propto f^{-1}$ and we
have assumed $C_{HJ}=1$ because the observations of HJ06 are deeper
than ours.  This posterior is effectively the distribution of $f$ for
objects that could have been detected by HJ06 and might be detected by
us after adjusting for our completeness, $C$.

Finally, we may use $P(f|E_{HJ})$ as a Bayesian prior for our TALCS
study where we find $n$ MBCs in $M$ asteroids:
\begin{eqnarray}
\label{eq:pfetalcs1}
P(f|E_{\rm TALCS}) &=& {   (Cf)^n (1-Cf)^{M-n-1}  P(f|E_{HJ}) \over
                      \int_0^1  P(f|E_{HJ}) (Cf)^n (1-Cf)^{M-n-1} \,df  }\\
\label{eq:pfetalcs2}
                 &=& {   (Cf)^n (1-Cf)^{M-n-1} (1-f)^{M_{HJ}}   \over
                      \int_0^1     (Cf)^n (1-Cf)^{M-n-1}  (1-f)^{M_{HJ}} \,df  }
\end{eqnarray}
To compute uncertainties it is necessary to integrate $P(f|E_{\rm
  TALCS})$ up to the desired confidence boundary.  For the case $C=1$,
Equation \ref{eq:pfetalcs1} simplifies to a ratio of incomplete beta
functions.

Although the Bayesian approach addresses the problem of inconsistency
with prior knowledge it does not resolve the binning difficulty.  As
smaller bins of new data are considered the recovered MBC fraction
defaults to the Bayesian prior.  In fact, the absence of MBCs in
neighboring bins should provide information on the number expected in
a particular bin because there is no reason to believe that the bins
are completely independent.  For instance, we do not genuinely believe
that the MBC fraction for semimajor axis $a\in[2.1,2.2]$ is given by
$f\sim{M_{\rm HJ}}^{-2}$ when we oberved zero MBCs out of a thousand
asteroids at other values of $a$. A correct treatment would require
assigning a prior probability to the independence of the bins.

\subsection{Conclusion} 

Because of the problems discussed above, one cannot view the formal
bound $f_{90}$ as a simple ``betting'' confidence and any
interpretation of the limits must be in light of the caveats of this
Appendix.  The simplest interpretation may be the most reliable: if we
use the HJ06 result as a Bayesian prior and consider our sample as a
whole, we arrive at
\begin{equation}
P(f|E_{\rm TALCS}) = (M_{\rm HJ}+M_{\rm TALCS})\times(1-f)^{M_{\rm HJ}+M_{\rm TALCS} -1 }
\end{equation}
which is identical to a single combined experiment that discovered one MBC.

\end{appendix}

\end{document}